\documentstyle[11pt,epsf,epsfig,psfig]{article}

\title{\bf CENTRAL RAPIDITY DENSITIES OF CHARGED PARTICLES
AT RHIC AND LHC}
\author{N. Armesto\thanks{E-mail: fa1arpen@uco.es.}\\
{Departamento de F\'{\i}sica, M\'odulo C2,
Planta baja, Campus de Rabanales,}\\
{\it Universidad de C\'ordoba, E-14071 C\'ordoba, Spain}\\
and\\
C. Pajares\thanks{E-mail:
pajares@gaes.usc.es.}\\
{\it Departamento de F\'{\i}sica de
Part\'{\i}culas, Universidade de Santiago de Compostela,}\\
{\it E-15706 Santiago de Compostela, Spain}\\
\\
\\
}
\date{February 2000}
\def\beq{\begin{equation}}
\def\eeq{\end{equation}}

\begin{document}
\maketitle
\medskip
\centerline{\bf Abstract}

Predictions on central rapidity
densities of charged particles at energies of the Relativistic Heavy Ion
Collider and the Large Hadron Collider, for central collisions between
the largest nuclei that will be available at these accelerators,
are reviewed.
Differences among the results of
the existing models are discussed in relation with
their underlying physical basis and with the possibilities to discriminate
them.

\vfill
\rightline{US-FT/2-00}
\rightline{UCOFIS 1/00}
\rightline{hep-ph/0002163}

\newpage

\section{Introduction}
\noindent
In the last years many experimental and theoretical efforts have been
devoted to search Quark Gluon Plasma (QGP) and/or collective
effects\footnote{Collective effects are those considered to explain the
event, which go beyond
the
superposition of elementary nucleon-nucleon collisions.}~
in heavy
ion collisions \cite{hwa,books,qm991,qm992,brown,c1}, as a tool to study the
nonperturbative aspects of Quantum Chromodynamics (QCD). To achieve this
goal several signatures of the phase transition(s) from confined to
deconfined quarks and gluons and of chiral symmetry restoration have
been proposed.

The finding of three of the proposed \cite{ms,rm,raf} signals
at the Super Proton
Synchrotron (SPS) at CERN has originated great excitement and debate in
the scientific community in this field. In fact, an abnormal suppression
of J/$\psi$ \cite{na50}
and a strong enhancement of strange baryons and antibaryons
\cite{strange}
have been observed in central\footnote{By central collision we mean
a head-on one,
in which most of the matter of the lightest nucleus
participates. In practice different criteria are used,
both theoretically (upper bound in impact parameter, minimum number of
participant wounded nucleons,$\dots$) and experimentally (percentage of
the cross section, lower bound in the number of charged
particles,$\dots$).}
~ Pb-Pb collisions, compared with those
measured in collisions between lighter projectiles and targets. Also an
enhancement in the dilepton spectrum for dilepton masses below 0.8
GeV/c$^2$ has been seen in Pb-Au collisions \cite{dilep}.
Whether or not these three experimental observations are really an
unambiguous proof of the existence of a QGP is still an
open question \cite{bo,wong,kns,abfp,acf,ggbcm,vgw,cfs,lkb},
due to their possible explanation using more
conventional, but still interesting, physics.
In any case, it is expected that the forthcoming heavy ion experiments
in the Relativistic Heavy Ion Collider (RHIC) at BNL and the Large
Hadron Collider (LHC) at CERN
will clarify definitively the
point\footnote{RHIC and LHC will provide center of mass
energies of 200 GeV and 5.5 TeV per nucleon respectively, to be compared
with $\sim 20$ GeV per nucleon at the SPS.}~.

In spite of the work done these last years there are many fundamental
aspects of the physics of heavy ions at high energies which are not
clear at all. Fundamental questions like, e.g.:
\begin{itemize}
\item Is particle rapidity density proportional to the number of participant
nucleons or to the number of elementary nucleon-nucleon collisions?
\item Which is the physical explanation of the SPS particle correlation
data?
\item How large will particle multiplicities be at RHIC and LHC?
\end{itemize}
are answered in a very different way by several models, all of them
claiming to agree with
the existing experimental data.

\begin{figure}[htbp]
\vspace*{13pt}
\begin{center}
\epsfig{file=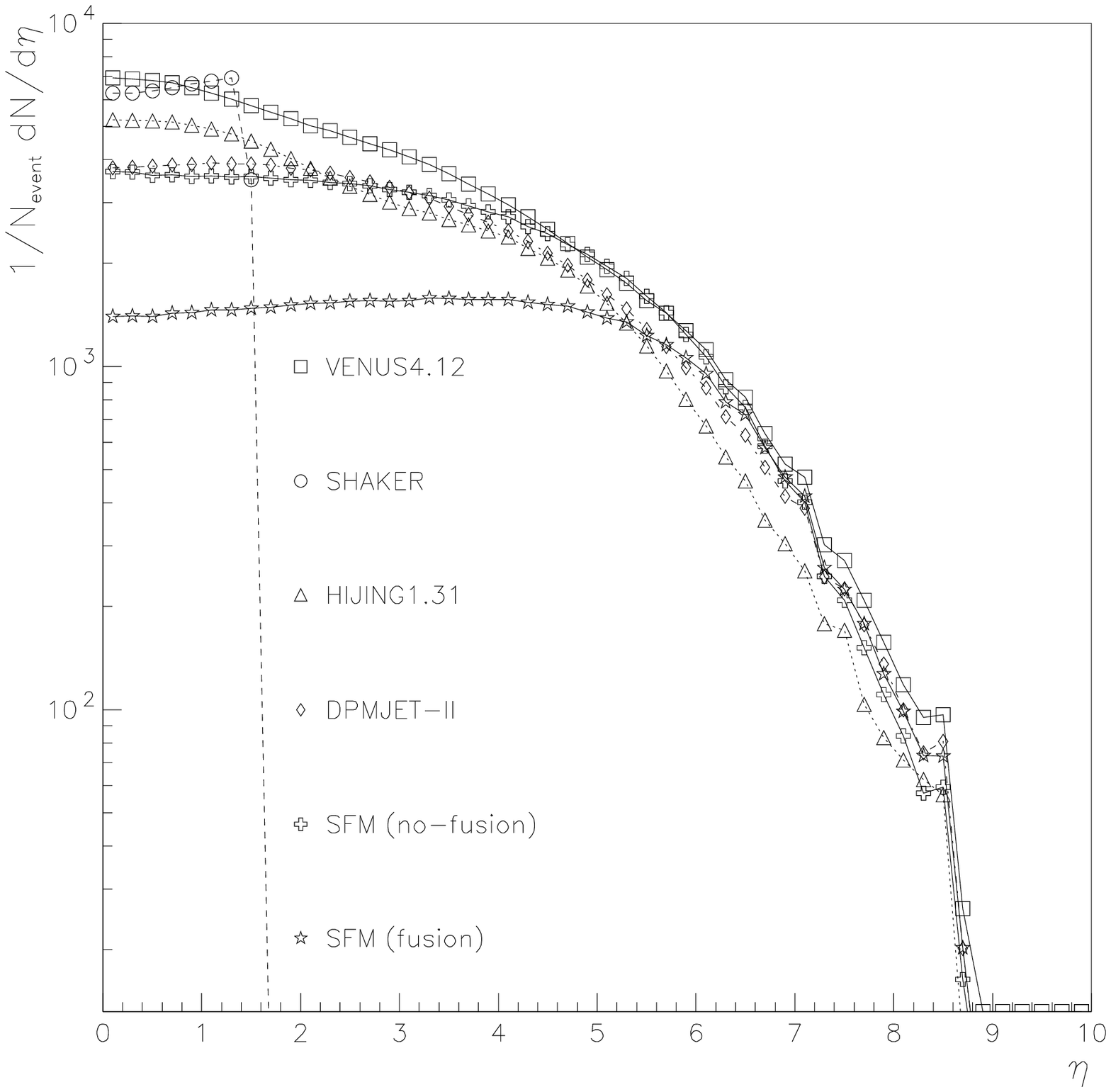,width=9.cm}
\end{center}
\vspace*{13pt}
\caption{Predictions from different models for the charged pseudorapidity
density in central ($b\leq 3$ fm) Pb-Pb collisions at the LHC, taken from
\protect{\cite{alice}}.}
\label{fig1}
\end{figure}

Referring to the last question, in Fig. \ref{fig1} it is shown the
pseudorapidity distribution of charged particles from different models for
central Pb-Pb collisions at a beam energy of 3 TeV per nucleon; this
plot has been taken
from the ALICE\footnote{ALICE (A Large Ion Collider Experiment) is the
approved detector at the LHC fully dedicated to Heavy Ion Physics.}~ Technical
Proposal \cite{alice} done by the ALICE Event
Generator Pool in December
1995. The results according to Monte Carlo codes of several models show
large differences at central pseudorapidity. Indeed, between the String
Fusion Model (SFM) \cite{sfm} and the VENUS \cite{venus}  or SHAKER
\cite{shaker} codes there is a factor larger than 4 at $\eta=0$, while
the difference in the fragmentation regions ($|\eta|\geq  5$) is smaller.
At RHIC energy\footnote{Un updated set of predictions for RHIC can be
found in \protect{\cite{lastcall}}.}~ the difference
is about a factor 2 - most models give
results in the range $700\div 1500$.

These uncertainties in one of the most elementary aspects of the
collision,
may leave us uncomfortable regarding the necessity to keep under control
the conventional physics of heavy ions to clearly distinguish the
signatures of QGP and/or collective effects in the proposed observables.
Needless to say, from the experimental point of view it is crucial to
know whether there will be 2000 or 8000 charged particles per unit
rapidity in central Pb-Pb collisions at the LHC for the
design of the detectors. For these reasons we review in this paper
charged particle
central rapidity density predictions of different models for central
collisions between the largest nuclei that will be available at RHIC and
LHC, discussing the
origin of the differences among the results.

According to their origin, models
can be classified into three categories. On the one hand, some models
like Dual Parton Model (DPM) \cite{dpm,ckt,cpr}, its Monte Carlo implementation
DPMJET \cite{dpmjet1,dpmjet2}, Quark-Gluon String Model (QGSM) \cite{qgsm},
FRITIOF \cite{fritiof},
SFM \cite{sfm,abctg,us}, Relativistic Quantum Molecular Dynamics (RQMD)
\cite{rqmd1,rqmd2,rqmd3}, Ultrarelativistic Quantum Molecular Dynamics
(UrQMD) \cite{urqmd1,urqmd2}, VENUS \cite{venus} or its new version
NEXUS \cite{hladik},
and LUCIAE \cite{luciae} mainly pay attention to the
soft part of the collision (there is no need of a hard perturbative part
at SPS energies). The hard part in some of these models is included
adding to the elementary soft cross section the jet one, as an input for
the eikonalized cross section.

On the contrary, other models like the Heavy-Ion Jet Interaction
Generator (HIJING) \cite{hijing}, Eskola {\it
et al.} \cite{eskola}, and Geiger and M\"uller \cite{geiger} are mainly
focused to the hard part. They compute the number of minijets or partons
with transverse momentum larger than a given $p_{\perp 0}\geq 1\div 2$
GeV/c. These hard partons are taken \cite{hardinit}
as the starting point of an
evolution and expansion previous to hadronization (for discussions on
this point see
for example \cite{kajan,mueller}). A soft part,
extracted from the SPS data, is added with an energy dependence taken
from some model.

A third kind of models are the statistical and thermodynamical ones
\cite{qm992,brown}. In these models the main predictions refer to ratios
between different kind of particles and not to absolute values of each
kind. Usually to get absolute rapidity densities the volume at freeze-out
has to be specified. The volumes used are $\sim 3600$ and $\sim 14400$
fm$^3$, giving
charged particle densities at midrapidity of 1200 and 8000 at RHIC and
LHC energies respectively\footnote{The value of 8000 charged particles
per unit rapidity at $y=0$ was the preferred value for many models before
1995. Indeed only the SFM gave values close to 2500. Now, by different
although probably related reasons, several models
have lowered their predictions to values close to the SFM one.}~.

The plan of this review will be the following: After this Introduction, in
the next Section the DPM and DPMJET Monte Carlo code will be discussed
in some detail, introducing several concepts which will also be used in
the other models. In Sections 3, 4, 5 and 6
the SFM, RQMD, HIJING and Perturbative QCD (PQCD) and Hydrodynamical
models respectively, together with their
predictions for charged particle densities at midrapidity,
will be briefly reviewed. Other models will be discussed in Section
7. Afterwards, in Section 8 we will argue on
percolation in heavy ion collisions, and in Section 9
possible implications for Cosmic Ray Physics will be commented.
In the last Section
the different results will be compared and some discussions
presented.

\section{The Dual Parton Model and the DPMJET Monte Carlo code}
\begin{figure}[htbp]
\vspace*{13pt}
\begin{center}
\epsfig{file=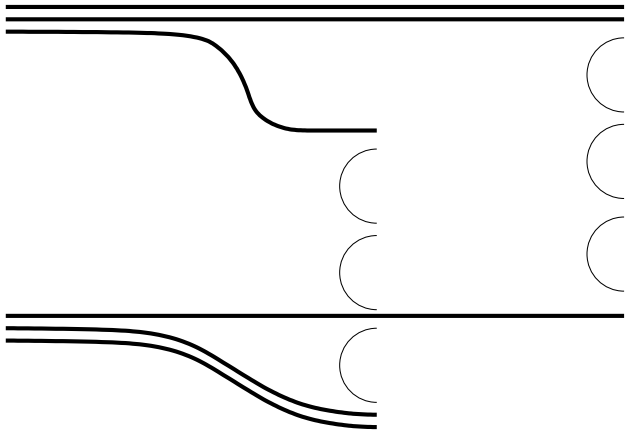,width=12.5cm}
\end{center}
\vspace*{13pt}
\caption{Lowest order contribution to p-p scattering at high energies
in the DPM. Strings are stretched between valence constituents of the
protons and hadronize by means of $q\overline{q}$ pair production.}
\label{fig2}
\end{figure}

\noindent
The DPM \cite{dpm} is a dynamical model for low $p_\perp$ hadronic and nuclear
interactions, based on the large $N$ expansion of QCD with $N_c/N_f$
fixed \cite{gabriele}. The dominant lowest order configuration in p-p scattering
at high energy consists in the production of two strings between valence
constituents, of type $(qq)_v-q_v$, see Fig. \ref{fig2}.

\begin{figure}[htbp]
\vspace*{13pt}
\begin{center}
\epsfig{file=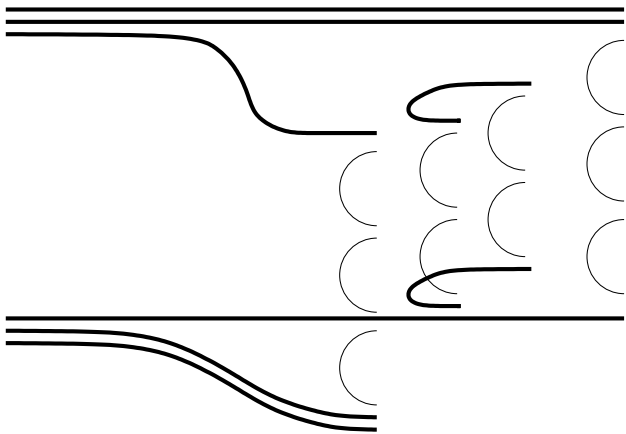,width=12.5cm}
\end{center}
\vspace*{13pt}
\caption{First higher  order contribution to p-p scattering at high energies
in the DPM. Besides the contribution shown in Fig. \protect{\ref{fig2}},
now strings stretched between sea constituents of the
protons appear.}
\label{fig3}
\end{figure}

There are also more complicated terms, corresponding to higher order
diagrams in the large $N$ expansion, involving 4, 6,$\dots$ strings. These
extra strings are of the type $q_s-\overline{q}_s$, with sea quarks and
antiquarks at their ends (Fig. \ref{fig3}). These configurations
correspond to multiple inelastic scattering in the $S$-matrix approach,
the number of strings being equal to twice the number of inelastic
collisions. The contribution of each configuration to the cross section
is determined using the generalized eikonal approach (see below)
or the perturbative
Reggeon calculus \cite{reggeon} in hadron-hadron collisions and the
Glauber-Gribov
model \cite{glauber,gribov} in collisions involving nuclei.

For A-A collisions, the rapidity distribution of secondaries is given by
\cite{ckt,cpr}
\begin{equation}
\frac{dN^{\rm AA}}{dy} = \overline{n}_{\rm A}\left[ N^{(qq)_v^{{\rm
A}_p}-q_v^{{\rm A}_t}}(y)+N^{q_v^{{\rm 
A}_p}-(qq)_v^{{\rm A}_t}}(y)\right] + 2\ (\overline{n}-\overline{n}_{\rm
A})\  N^{q_s
-\overline{q}_s}(y),
\label{eq1}
\end{equation}
where $N(y)$ are the rapidity distributions of produced particles in the
individual strings stretched between the projectile ($p$) and target
($t$) nuclei, $\overline{n}_{\rm A}$ is the average number of wounded
nucleons of A and $\overline{n}$ is the average number of
nucleon-nucleon collisions. Both $\overline{n}_{\rm A}$ and
$\overline{n}$ are computed \cite{bbz} in the Glauber model. For
instance, for minimum bias collisions
\begin{equation}
\overline{n}=\frac{{\rm A}^2\sigma_{\rm NN}}{\sigma_{\rm AA}}\simeq  \frac{{\rm
A}^2\sigma_{\rm NN}}{\pi (2 R_{\rm A})^2}\approx \frac{{\rm A}^{4/3}}{4}
\ ,
\label{eq2}
\end{equation}
with $\sigma_{\rm NN}$ and $\sigma_{\rm AA}$ the nucleon-nucleon and
nucleus-nucleus cross sections respectively;
for central collisions,
\begin{equation}
\sigma_{\rm AA}\simeq \pi R_{\rm A}^2 \ ,\ \ \overline{n}
\approx {\rm
A}^{4/3}.
\label{eq3}
\end{equation}

If all strings would have the same plateau height (i.e. the same value
of $N(0)$), $dN^{\rm AA}/dy|_{y=0}$ would increase like A$^{4/3}$.
However at present energies the plateau height of the
$q_s-\overline{q}_s$ strings is much smaller than that of the
$(qq)_v-q_v$ ones, and the first term in (\ref{eq1}) dominates. One
obtain in this way the result of the Wounded Nucleon Model (WNM)
\cite{bialas}. At SPS energies only for central collisions some
departure of the law \cite{cpr}
\begin{equation}
\frac{dN^{\rm AA}}{dy} \propto \overline{n}_{\rm A}
\label{eq4}
\end{equation}
is expected, and indeed has been seen in the experimental data
\cite{na49qm99,wa98}.

At higher energies the contribution of the sea strings becomes
increasingly important, not only because their plateau height gets
higher but also due to the need to introduce multistring configuration
in each nucleon-nucleon collision. If the average number of strings in
each nucleon-nucleon collision is $2\overline{k}$ (this number can be
computed in the generalized eikonal model), the total number of strings
is $2\overline{k}\overline{n}$ and (\ref{eq1}) is changed into
\begin{eqnarray}
\frac{dN^{\rm AA}}{dy} &=& \overline{n}_{\rm A}\left[ N^{(qq)_v^{{\rm
A}_p}-q_v^{{\rm A}_t}}(y)+N^{q_v^{{\rm 
A}_p}-(qq)_v^{{\rm A}_t}}(y)+(2\overline{k}-2)
N^{q_s
-\overline{q}_s}(y)\right]\nonumber \\
& +& 2\ \overline{k}\ (\overline{n}-\overline{n}_{\rm
A})\  N^{q_s
-\overline{q}_s}(y).
\label{eq5}
\end{eqnarray}

The hadronic spectra of the individual strings $N(y)$ is obtained from a
convolution of momentum distribution functions and fragmentation
functions \cite{dpm}. Both functions can be determined to a large extent
from known Regge trajectories.

For RHIC and LHC energies $\overline{k}$ $\sim 2$ and 3 respectively.
Using these values in (\ref{eq5}) it is obtained \cite{cmt}:
\begin{eqnarray}
\left.\frac{dN^{\rm SS}}{dy}\right|_{y=0}=170, && \left.\frac{dN^{\rm
PbPb}}{dy}\right|_{y=0}=1890\ \ \mbox{ at }\sqrt{s}=200 \mbox{
GeV per nucleon},\nonumber \\
\left.\frac{dN^{\rm SS}}{dy}\right|_{y=0}=500, && \left.\frac{dN^{\rm 
PbPb}}{dy}\right|_{y=0}=7900\ \ \mbox{ at }\sqrt{s}=7 \mbox{ 
TeV per nucleon},
\label{eq6}
\end{eqnarray}
for charged particles in central ($\overline{n}_{\rm A}>28$ in S-S and
200 in Pb-Pb, corresponding to $b\simeq 0$)
A-A collisions (in \cite{ckat} a value
of 8500 for Pb-Pb, $b< 3$ fm, at $\sqrt{s}=6$ TeV per nucleon is given).

In these results no semihard collisions were taken into account. The
inclusion of this kind of collisions cannot modify significantly the
numbers in (\ref{eq6}) since the total number of strings is constrained
by unitarity. The fact that some of the $q_s-\overline{q}_s$ strings can
be the result of a semihard gluon-gluon interaction, will affect the
$p_\perp$ distribution of the produced particles. However, average
multiplicities are practically unchanged if one neglects changes from
energy-momentum
conservation due to the larger $p_\perp$ in the semihard contribution.

Other effect not taken into account in these first estimations done in 1991
\cite{cmt} is shadowing corrections, which can be of importance at RHIC
and LHC energies. The physical origin of shadowing
corrections\footnote{For a discussion on the relation between
unitarity, parton saturation and shadowing see for example
\protect{\cite{mueller}}.}~ can be traced back to the diffe\-ren\-ce between
the space-time
picture of the interaction
in the Glauber model \cite{glauber} and in Glauber-Gribov field
theory \cite{gribov}.
In Glauber
we have successive collisions of
the incident hadron to explain
multiple scattering in hadron-nucleus interactions,
while in Gribov theory simultaneous collisions of
different projectile constituents with nucleons in the target nuclei
are considered.
Nevertheless, the h-A scattering amplitude can be written as a sum of
multiple scattering diagrams with elastic intermediate states, which
have the same expressions in both cases. In addition to these diagrams
there are other ones which contain, as intermediate states,
all possible diffractive excitations
of the projectile hadron, whose influence at SPS energies is
small. The size of the high mass
excitations of the initial hadron is controlled by the triple Pomeron
coupling. The value of this coupling, determined from soft diffraction
experimental data, allows to describe hard diffraction measured at the
Hadron-Electron Ring Accelerator (HERA) at DESY and also the size of the
shadowing effects in the nuclear structure functions at small $x$
\cite{ckmpt}. These considerations imply a reduction of particle
densities at midrapidity of a factor 2 at RHIC and 3 at the
LHC \cite{ckat}.

This shadowing can be alternatively seen as a way of
introducing the interaction among strings, see next Section. In (\ref{eq1}) and
(\ref{eq5}) it is assumed that strings fragment independently one from
each other. As the number of strings grows with the energy of the
collision, with the size of the projectile or the target and with the
degree of centrality of the collision, interaction of strings is
expected at very high energies in central heavy ion collisions. This
approach is equivalent to take into account the triple Pomeron coupling,
whose effects on $dN/dy$ are very small at SPS energies and become large
at RHIC and LHC.

In order to include the hard part in the DPM, the eikonal depending on
impact parameter $b$ and energy is divided in
a sum of soft plus hard pieces \cite{hard,zjil},
\begin{equation}
\chi(b^2, s)= \chi_s(b^2, s)+\chi_h(b^2, s),
\label{eq7}
\end{equation}
normalized to the corresponding elementary cross sections,
\begin{equation}
\int d^2b \ 2\chi_i(b^2, s)=\sigma^0_i\ ,\ \ i=s,h;
\label{eq8}
\end{equation}
in terms of the eikonal, the inelastic cross section for the collision
is
\begin{equation}
\sigma_{in}=\int d^2b \ \left[1-e^{-2\chi(b^2, s)}\right].
\label{eq9}
\end{equation}
The soft eikonal is parametrized as
\begin{equation}
\chi_s(b^2, s)= \frac{\sigma^0_s}{8\pi\left[c+\alpha^\prime
\log{(s/s_0)}\right]}\ \exp{\left(-\frac{b^2}{4\left[c+\alpha^\prime
\log{(s/s_0)}\right]}\right)}
\label{eq10}
\end{equation}
and the hard one as
\begin{equation}
\chi_h(b^2, s)= \frac{\sigma^0_h}{8\pi d}
\exp{\left(-\frac{b^2}{4d}
\right)}\ .
\label{eq11}
\end{equation}
The soft input is a soft Pomeron with a linear trajectory,
$\alpha_s(t)=1+\Delta_s+\alpha^\prime t$,
\begin{equation}
\sigma^0_s=g^2 s^\Delta_s,
\label{eq12}
\end{equation}
and the hard cross section $\sigma^0_h$ is calculated from PQCD using a
lower $p_\perp$ cut-off and conventional structure functions. Unitarity
of the cross section is explicit in (\ref{eq9}), which can be expanded
as
\begin{equation}
\sigma_{in}=\int d^2b \ \sum_{l_c+m_c\geq 1} \sigma(l_c,m_c,b^2,s),
\label{eq13}
\end{equation}
the sum running over $l_c$ soft elementary collisions and $m_c$ hard ones.

DPMJET is a Monte Carlo code for sampling hadron-hadron, hadron-nucleus,
nucleus-nucleus, lepton-hadron and lepton-nucleus collisions at
accelerator and cosmic ray energies \cite{dpmjet1}. It uses the DPM for
hadronic and nuclear interactions, the hard part being
simulated using PYTHIA \cite{pythia} and, in its latest version
DPMJET-II.5 \cite{dpmjet2}, one
of the most recent sets of
parton distribution functions, GRV-LO-98
\cite{grv}. The code includes intranuclear
cascade processes of the created secondaries with formation time
considerations, and also nuclear evaporation and fragmentation of the
residual nucleus.

\begin{figure}[htbp]
\vspace*{13pt}
\begin{center}
\epsfig{file=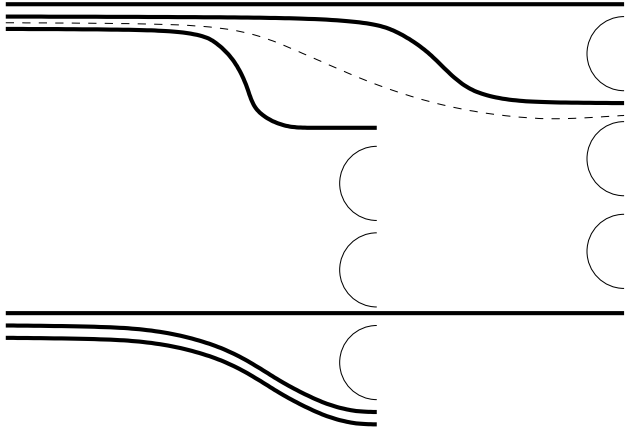,width=12.5cm}
\end{center}
\vspace*{13pt}
\caption{The popcorn mechanism of diquark breaking.}
\label{fig4}
\end{figure}

\begin{figure}[htbp]
\vspace*{13pt}
\begin{center}
\epsfig{file=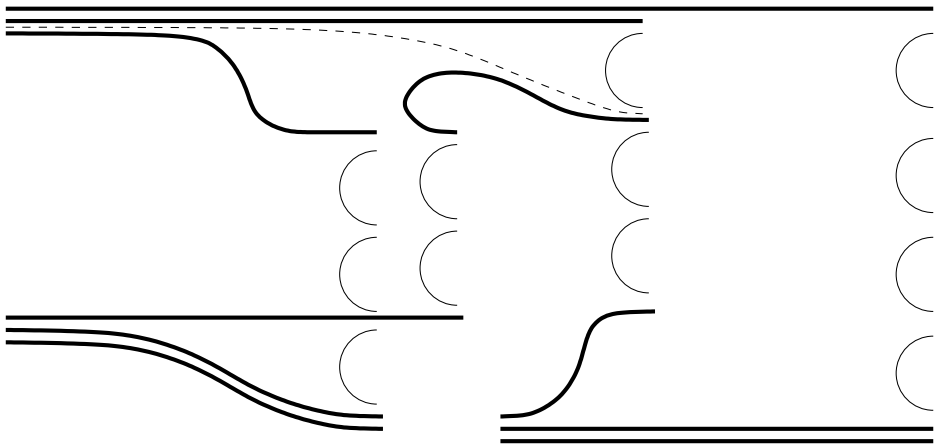,width=12.5cm}
\end{center}
\vspace*{13pt}
\caption{One example of proposed diagram for diquark breaking in
nucleon-nucleus collisions.}
\label{fig5}
\end{figure}

In the first versions of the code, in addition to diagrams where the
valence diquarks at the end of one string fragment into hadrons
preserving the diquark, diquark breaking was allowed. This is the
so-called popcorn mechanism, see Fig. \ref{fig4}. However the mechanism
is not enough to explain the large baryon stopping observed in A-B
collisions. For this reason in the DPM
new diagrams \cite{vgw,cfs,ck,vg} for diquark breaking, like the
one in Fig. \ref{fig5}, have been proposed and discussed. In both Figs.
\ref{fig4} and \ref{fig5} the dashed line is the string junction: at
large $N_c$ a baryon can be pictured \cite{rv} as made out of three
valence quarks together with three strings which join in the
string junction\footnote{Using some supergravity solution and the
recently conjectured duality between gauge and string theory the large
$N_c$ baryon wave function has been constructed from $N_c$ strings
connected via a junction \protect{\cite{gross}}.}~. These diagrams,
included in DPMJET-II.5,  shift
the baryon spectrum to the central rapidity region and also produce an
enhancement of strange baryons and antibaryons.

In the code the presence of diquarks and antidiquarks at sea string ends
is also included. This increases baryon and antibaryon rapidity
densities and, due to energy-momentum conservation, reduces that of
pions.

The results of DPMJET-II.5 for charged particles in
central Pb-Pb collisions at RHIC (3 \% more central events)
and LHC (4 \% more central events)
are \cite{dpmjet2}
\begin{equation}
\left.\frac{dN}{dy}\right|^{\rm RHIC}_{y=0}=1280,\ \
\left.\frac{dN}{dy}\right|^{\rm LHC}_{y=0}=2800.
\label{eq14}
\end{equation}
These values agree with the ones computed in the DPM \cite{ckat}, see
above. The previous version of the code \cite{dpmjet1} gives a higher
value at LHC, $dN_{ch}/d\eta|_{\eta=0}=3700$ \cite{alice}
(although for $\sqrt{s}=6$ TeV
per nucleon and $b\leq 3$ fm).
This reduction is due to the inclusion of new diagrams and
to the energy-momentum conservation consequences of the inclusion of
$(qq)_s-(\overline{q}\overline{q})_s$ strings. Notice than the value
obtained in the code is much smaller than that obtained in the DPM using
(\ref{eq5}). This fact is essentially due to energy-momentum
conservation, which prevents some of the
($\overline{n}$,$\overline{n}_{\rm A}$,$\overline{k}$)
configurations to take place.

\section{The String Fusion Model}
\noindent
The SFM \cite{sfm,abctg,us} is based on QGSM, a model which is quite similar
to DPM with only minor differences. The main ingredient added in the SFM
is the fusion of strings \cite{bp,ranftmerino}.
The basic idea is that strings fuse as soon as their transverse position
come within a certain interaction area, of the order of the string
proper transverse dimension as dictated by its mean $p_\perp$. In a
Monte Carlo approach, such a picture can be realized by assuming that
strings fuse as soon as partons which act as their sources, have their
transverse positions close enough. In this language the fusion
probability is determined by the parton transverse dimension, that is,
by the parton-parton cross section. Energy conservation can be taken
into account by distributing the available energy among these active
partons, as it has always been done in string models \cite{dpm,qgsm}.
Then the emerging strings occupy different intervals in rapidity space,
determined by the energy-momentum of their sources. The fusion of
strings may only take place when their rapidity intervals overlap. In
particular, for two pairs of partons from the projectile and target with
rapidities $y_1$, $y_2$ and $y_1^\prime$, $y_2^\prime$ respectively, the
two corresponding strings fuse in the interval $[{\rm
max}\{y_1^\prime,y_2^\prime\},{\rm min}\{y_1,y_2\}]$. If this interval
becomes small the resulting object will have will have its total energy
of the order of a typical hadron mass and, as with ordinary strings, is
no more a string but rather an observed hadron. The exact value of the
minimal string energy and thus of its minimal rapidity length is taken
the same as for ordinary strings.

The color and flavor properties of the formed strings follow from the
properties of their ancestor strings. The fusion of several
quark-antiquark $q-\overline{q}$ strings produces a $Q-\overline{Q}$
complex
with color $Q$ (quadratic Casimir operator of the representation
$Q^2$), which is determined by the SU(3) color composition
laws. For example, the fusion of two $q-\overline{q}$ triplet strings
produces a
$[\overline{3}]$ string (that is, a diquark-antiquark string) with
probability 1/3 and a $[6]$ string with probability 2/3 ($[3]\otimes[3]
=[6]
\oplus
[\overline{3}]$). On the other hand, if two triplet strings with opposite
color flux directions fuse (a quark $[3]$ state fuses with a
$[\overline{3}]$ antiquark state), either colorless states at the end of
the new string or a $[8]$ string are formed with probabilities 1/9 and
8/9 respectively ($[3]\otimes [\overline{3}]= [1]\oplus [8]$). The
flavor of the fused string ends is evidently composed of the flavor of
the partons sitting there. As a result of string fusion, we thus obtain
strings with arbitrarily large color and differently flavored ends, in
accordance with the probability to create the color $Q$ from several
(anti)quarks. Crude characteristics of hadron interactions depend only
on the fact that the total number of strings of whatever color in a
given transverse area becomes limited because of their fusion. In other
words, string density cannot grow infinitely but is bounded from above
\cite{bp}. More detailed properties of hadron spectra require knowledge
of a particular manner in which the new fused strings decay into
hadrons.

In all color string models, it is assumed that the homogeneous color
field corresponding to the strings creates pairs of colored partons,
which neutralize this field and provide for its subsequent decay. The
basic formula which  describes the probability of such a process is
taken in the spirit of the famous Schwinger expression for the
probability to create an electron-positron pair in a constant
electromagnetic field \cite{schwinger,casher,biro}. With a constant
color field which originates from two opposite color charges $\vec{Q}$
and $-\vec{Q}$ (8-vectors in SU(3)), the probability rate to create a
pair of partons with color charges $\vec{C}$ and $-\vec{C}$, flavor $f$
and mass $M_f$ for unit string length is assumed to be given by
\begin{equation}
\frac{d\omega(\vec{Q},\vec{C})}{d^2 p_\perp} \propto A_t
\ (k{Q}{C})^2\ 
\exp{\left(-\frac{M_\perp^2}{k{Q}{C}}\right)}\ .
\label{eq15}
\end{equation}
The parameter $A_t$ has the meaning of the string transverse area and
$M_\perp$ is the transverse mass. $k$
is proportional to the string tension $\kappa$,
\begin{equation}
\kappa = \frac{\pi k Q^2}{2}\ .
\label{eq16}
\end{equation}

In \cite{biro,andersson} the strings of high color, denoted as color
ropes, break as a result of successive production of $q\overline{q}$
pairs, which gradually neutralize the color flux of the string until it
breaks. In SFM, however, the process considered is that of creation of a
pair of parton complexes with color $\vec{Q}$ equal to that of the ends
of the string. This is the main contribution to the
breaking of the string for low
values of $Q$ \cite{sfm}. As, in the Monte Carlo code, fusion of strings
is taken into account in an effective way and only fusion of two strings
is allowed, the mechanism of string breaking for high color strings is
the one just mentioned. It is also assumed that $[3]$, $[\overline{3}]$,
$[6]$ and $[8]$
strings have the same transverse area\footnote{This assumption is a very
strong one; other possibilities will be discussed in Section 8
in relation to
percolation of strings \cite{abfp,nardi,bpr,abfpp}.}~. The string
tension is proportional to $Q^2$,
\begin{equation}
Q^2_{[3]}=4/3=Q^2_{[\overline{3}]},\ \ Q^2_{[8]}=3,\ \ Q^2_{[6]}=10/3.
\label{eq17}
\end{equation}
So, approximately $\kappa_{[8]} \sim \kappa_{[6]} \sim 2.5\ 
\kappa_{[3]}=2.5\ \kappa_{[\overline{3}]}$.

As can be inferred from what has been presented above,
fusion of strings leads to an enhancement of baryon and antibaryon
production, due to the
possibility of having $(qq)$ and $(\overline{q}\overline{q})$ at the end of
the fused strings. In DPM a similar mechanism is introduced considering
the possibility of diquarks in the sea, as mentioned in the previous
Section. In addition to this mechanism there is another source of baryon
enhancement, due to the larger tension of fused strings (\ref{eq17})
which, through (\ref{eq16}) and (\ref{eq15}), implies a more efficient
production of heavy quarks and diquarks, and of higher $p_\perp$.
Therefore, it is also expected
heavy flavor enhancement and some increase of transverse momentum.

Another important consequences of string fusion is the possibility of
producing particles in collisions involving nuclei, outside the
nucleon-nucleon kinematical region, the so-called cumulative effect
(part of these effect is usually addressed to the Fermi motion of
nucleons inside nuclei). In
fact the resulting fused string has an energy-momentum corresponding to
the sum of the energy-momenta of its ancestor strings, which can be
larger than the energy-momentum available in an isolated nucleon-nucleon
collision \cite{cosmic1}.

In the SFM code, the nuclear parton wave function is taken as a
convolution of the parton distribution in a nucleon with the nucleon
distribution in a nucleus. In this way, hadrons and nuclei are treated
in a similar way, different from what DPMJET does.
Also, in a previous version of SFM \cite{sfm} most of the computations
where done at SPS energies and no hard part was considered. This part is
now introduced in the code in a standard way \cite{us} and modifies the
central rapidity region at energies higher than those of SPS.

The probability of fusion of two strings is controlled by the
parton-parton cross section,
\begin{equation}
\sigma_p=2\pi r^2.
\label{eq18}
\end{equation}
Its numerical value is fixed to $\sigma_p\simeq 8$ mb,
in order to reproduce the
$\overline{\Lambda}$ enhancement seen in central S-S and S-Ag collisions
at SPS energies \cite{antilam}. This value, which means $r\sim 0.36$ fm,
has been obtained implementing in the code
fusion of only two strings and therefore
has to be considered as an effective one. The actual transverse size of
a string should be less, a more realistic one being $r\simeq 0.2\div
0.25$ fm \cite{abfp}, a value which agrees with other considerations
\cite{nardi,shuryak,bali}.

For the purpose of this review, the most important consequence of
fusion of strings is that it suppresses
total multiplicities, reducing the number of
pions in the central rapidity region, although the rapidity distribution
becomes larger at the extreme of the fragmentation regions. In the
predictions done with the previous version of the model \cite{sfm},
no hard part was included and the
charged particle densities at midrapidity for central ($b=0$) Au-Au
collisions were
\begin{eqnarray}
\left.\frac{dN}{dy}\right|_{y=0}&=&1000
\ \ \mbox{ at }\sqrt{s}=200 \mbox{
GeV per nucleon},\nonumber \\
\left.\frac{dN}{dy}\right|_{y=0}&=&1900
\ \ \mbox{ at }\sqrt{s}=6.3 \mbox{
TeV per nucleon}.
\label{eq19}
\end{eqnarray}
The corresponding values without considering fusion of strings were 1850
and 4000 respectively. A strong suppression of the central density is
produced (note the agreement with the values from DPMJET). Including the
hard part \cite{us}, the values for collisions of charged particle densities
corresponding to the 5\% more central events\footnote{In the model,
this translates
into $b\leq 3.2$ fm for Au-Au at RHIC and $b\leq 3.3$ fm for Pb-Pb at
LHC.}~ are
\begin{eqnarray}
\left.\frac{dN^{\rm AuAu}}{dy}\right|_{y=0}&=&910
\ \ \mbox{ at }\sqrt{s}=200 \mbox{
GeV per nucleon},\nonumber \\
\left.\frac{dN^{\rm PbPb}}{dy}\right|_{y=0}&=&3140
\ \ \mbox{ at }\sqrt{s}=5.5 \mbox{
TeV per nucleon},
\label{eq20}
\end{eqnarray}
and now the suppression due to string fusion is smaller (the
corresponding values without fusion are 1300 and 3690 respectively).
The reason for this
is that the strings coming from hard scatterings do not fuse in the
code. At LHC a large proportion of strings are hard ones and therefore
the relative size of suppression is smaller. The hard strings have a
size $\sim 1/p_\perp$ and indeed should interact and fuse, although with
smaller probability than the soft ones. Effects of overlapping of
strings
will be further discussed in Section 8.

\section{The Relativistic Quantum Molecular Dynamics Model}
\noindent
RQMD \cite{rqmd1,rqmd2,rqmd3} is a semiclassical microscopic approach
which combines classical propagation with stochastic interactions.
Strings and resonances can be excited in elementary collisions, their
fragmentation and decay leading to the production of particles. The nature
of the active degrees of freedom in RQMD depends on the relevant length
and time scales of the processes considered. In low energy collisions
(around 1 GeV per nucleon in the center of mass) RQMD reduces to solving
transport equations for a system of nucleons, other hadrons and
eventually resonances interacting in binary collisions and via mean
fields. At large beam energies ($> 10$ GeV per nucleon in the center of
mass) the description of a projectile hadron interacting in a medium (a
cold nucleus) as a sequence of separated hadron or resonance collisions
breaks down. A multiple collision series is formulated on the partonic
level, following the Glauber-Gribov picture. In RQMD these multiple
collisions correspond to strings formed between partons of the
projectile and target, including sea quarks and antiquarks. The string
excitation law $dP\propto dx^+/x^+$ is the same used in  FRITIOF
\cite{fritiof}. The decay of elementary color strings is done using
JETSET \cite{sjostrand}. Rescattering is included: four classes of
binary interactions, BB, BM, MM and $\overline{\rm B}$B (B denoting baryon, M
denoting meson) are considered.

One of the main ingredients of RQMD is the inclusion of interaction of
strings by means of formation of color ropes, see previous Section,
when there are overlapping strings. These ropes are chromoelectric flux
tubes whose sources are charge states in representations of color SU(3)
with dimension higher than the triplet one. They are equivalent to the
fused strings of SFM. As already mention, as a simplification in SFM
only fusion of two strings is considered, as an effective way to take
into account string interaction. In RQMD all possibilities are
considered. The breaking of these higher color strings proceeds through
successive production of $q\overline{q}$ pairs \cite{biro,andersson}
due to the Schwinger
mechanism.

As in the case of SFM, introduction of color ropes in RQMD leads to
heavy flavor and baryon and antibaryon enhancement. In version RQMD 2.3
the model reproduces the SPS rapidity distributions of h$^-$, K$^0$,
$\Lambda$, $\overline{\Lambda}$, $\Xi^-$ and $\overline{\Xi}^+$. It
slightly underestimates the yields of $\Omega^-$ and
$\overline{\Omega}^+$ (by less than a factor 2). Also it is able to
reproduce the $m_\perp$ spectrum of all these particles. Let us
mention that independent string models are not able to reproduce these
slopes.

The formation of color ropes leads to a strong suppression of central
rapidity distributions. The prediction of RQMD for central ($b=3$ fm)
Pb-Pb
collisions at RHIC is $dN/dy|_{y=0}\simeq 700$
\cite{lastcall,sorgeqm99}. This number is lower than the value of SFM,
910. The reason for that probably has to do with the strong fusion
probability used in RQMD. The effect of this strong string interaction
in some observables (like antibaryon enhancement) is compensated by
other processes (a large $\overline{\rm B}$B annihilation).

\section{The Heavy-Ion Jet Interaction Generator (HIJING)}
\noindent
In HIJING \cite{hijing} the soft contribution is modeled by
diquark-quark strings with gluon kinks induced by soft gluon radiation,
in a way very similar to the FRITIOF model \cite{fritiof}.
Since this model treats explicitly minijet physics through PQCD, the
transverse momentum in string kinks due to soft processes is limited
from above by a minijet scale $p_{\perp 0}=2$ GeV/c. Gluon radiation is
extended to the hard part of high $p_\perp$, which, together with the
use of momentum distribution functions for partons similar to those of
DPM, constitutes a difference with FRITIOF. Strings decay independently
by means of the JETSET \cite{sjostrand} routines. In addition to the low
$p_\perp < p_{\perp 0}$ gluon kinks, HIJING includes an extra low
$p_\perp$ transfer between the constituent quarks and diquarks at the
string ends. This extra $p_\perp$ is chosen to ensure an smooth
extrapolation in the $p_\perp$ distributions from the soft to the hard
regime.

Multiple minijet production with initial and final state radiation is
included along the lines of the PYTHIA model \cite{pythia}. First, the
cross section for hard parton scattering $\sigma_{jet}$ is computed in
PQCD at leading order (LO),
using a K-factor $\simeq 2$ to simulate higher order corrections.
The eikonal formalism, see Section 2,
is employed to calculate the number of minijets per inelastic
nucleon-nucleon collision. For A-A collisions at impact parameter
$b$
the total number of
jets is given by
\begin{equation}
N^{\rm AA}_{jet}(b)= \frac{{\rm A}^2\ T_{\rm AA}(b)}{\sigma_{\rm AA}(b)}\ 
\sigma_{jet}\ ,
\label{eq21}
\end{equation}
with
\begin{equation}
T_{\rm AA}(b)=\int d^2b^\prime \ T_{\rm A}(b-b^\prime) T_{\rm
A}(b^\prime),
\label{eq22}
\end{equation}
$T_{\rm A}(b)$ being the nuclear profile function normalized to 1 and
$\sigma_{\rm AA}(b)\simeq \int d^2b\  \{1-\exp{[-\sigma_{\rm NN} A^2 T_{\rm
A}(b)]}\}$ the A-A cross section \cite{cpr,bbz} for impact parameter $b$. For
central collisions $b=0$, $\sigma_{\rm AA}(b=0)\simeq 1$ and
\begin{equation}
N^{\rm AA}_{jet}(b)\approx \frac{{\rm A}^2}{\pi R_{\rm A}^2}
\ 
\sigma_{jet}\propto {\rm A}^{4/3}\ .
\label{eq23}
\end{equation}
Therefore, at high energies and for central nucleus-nucleus collisions
there will be many minijets.

In the model, jet quenching \cite{quenching} is included to enable the
study of the
dependence of moderate and high $p_\perp$ observables, on an assumed
energy loss per unit length $dE/dx$ of high energy partons traversing
the dense matter produced in the collision. The effect of including
jet quenching is a moderate enhancement of particle production in the
central rapidity region and to diminish the yield in the fragmentation
regions. Furthermore, in the last version of the model the mechanism of
string junction migration \cite{vgw} explained in Section 2 is included,
in order to shift baryons from fragmentation to central rapidity regions.

\begin{figure}[htbp]
\vspace*{13pt}
\begin{center}
\epsfig{file=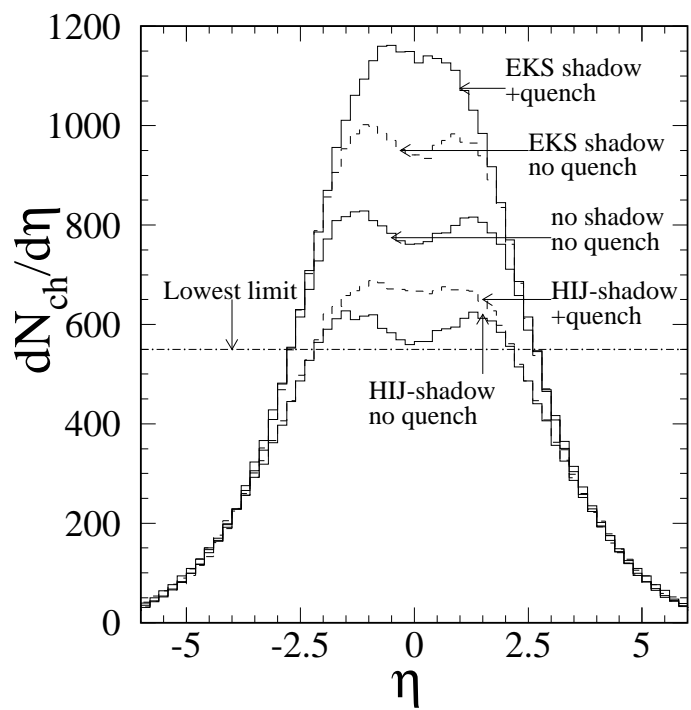,width=9.cm}
\end{center}
\vspace*{13pt}
\caption{Predictions for the charged pseudorapidity
density in central ($b<3$ fm) Au-Au
collisions at RHIC in the HIJING model, taken from
\protect{\cite{lastcall}}.}
\label{fig6}
\end{figure}

The results for charged densities at midrapidity in central ($b< 3$ fm)
Au-Au collisions at $\sqrt{s}=200$ GeV per nucleon are shown in Fig.
\ref{fig6}. The different curves refer to different versions of the
model with and without quenching and shadowing of the nucleon structure
functions in the nucleus \cite{hijing,eks98}. For LHC, HIJING predictions
\cite{hijing}
lie in the range $5000\div 7500$ depending on structure functions used and
quenching included or not.

\section{Perturbative Quantum Chromodynamics and Hydrodynamical models}
\noindent
It has been argued \cite{hardinit} that the initial state (the initial
distribution of partons) in a high
energy heavy ion collision could be computed using PQCD.
Several groups \cite{eskola,geiger} have developed models along this line. 
Concretely, Eskola {\it et al.} have computed \cite{eskola} charged
densities and transverse energies al midrapidities, using PQCD at some
given scale which is taken to be equal to a saturation scale, the
scale at which parton distributions stop their increasing at small $x$.

In ultrarelativistic heavy ion collisions the number of produced gluons
and quarks with $p_\perp$ greater than some cut-off $p_0$, $N_{\rm
AA}\left(b,p_0,\sqrt{s}\right)$,
 increases when $p_0$ decreases, when the size of
the nuclei increases or $b$ decreases, see (\ref{eq21}) and
(\ref{eq23}), and when $\sqrt{s}$ increases due to the small $x$
enhancement of parton distribution functions. Shadowing of nucleon
structure functions in nuclei will decrease $N_{\rm
AA}\left(b,p_0,\sqrt{s}\right)$,
but next-to-leading order (NLO) corrections will
increase it. At sufficient large cut-off $p_0 \gg \Lambda_{\rm QCD}$ the
system of produced gluons is dilute and usual perturbation theory is
applicable. However, at some transverse momentum $p_0=p_{sat}$ the
gluon and quark phase space density saturate and no further increase is
expected. In this case one may conjecture that evaluation of the number
of charged particles $N_{ch}$
and transverse energy $E_{T}$ using QCD
formulae at this saturation scale $p_{sat}$ gives a good estimate of the
total $N_{ch}$ and $E_{T}$ (partons with $p_\perp\gg p_{sat}$ are rare,
partons with $p_\perp\ll p_{sat}$ saturate and
contribute little to the total $E_T$).

\begin{figure}[htbp]
\vspace{5.3cm}
\hspace{0.5cm}
\epsfysize=12cm
\centerline{\epsffile{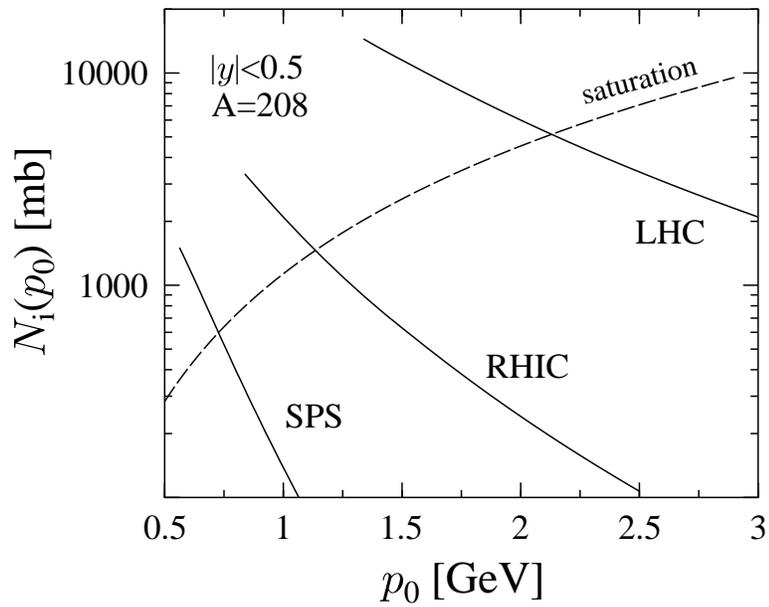}}
\vspace{-8.4cm}
\caption{$N_{\rm
AA}(b=0,p_0,\sqrt{s})$ with $p_T\ge p_0$ and $|y|<0.5$
for ${\rm A}=208$ as a function of
$p_0$ at $\sqrt{s}= 5500$
(LHC), 200 (RHIC) and 17 (SPS) GeV per nucleon, taken from
\protect{\cite{eskola}}.
The dashed curve is $p_0^2 R_{\rm
A}^2$.}
\label{fig7}
\end{figure}

In \cite{eskola}, first $N_{\rm
AA}\left(b=0,p_0,\sqrt{s}\right)$ for $|y|<0.5$
is computed using standard PQCD expressions at
LO. Nuclear effects on parton distribution functions are
implemented using the EKS98 parameterization \cite{eks98} of nuclear
corrections. To simulate NLO contributions, a K-factor ${\rm K}=2$ is
used. The scale in the PQCD calculation is fixed from considering that
at saturation $N_{\rm
AA}(b=0,p_{sat},\sqrt{s})$ partons, each one with transverse area
$\pi/p_{sat}^2$, fill the whole transverse area $\pi R_{\rm A}^2$,
\begin{equation}
N_{\rm AA}\left(b=0,p_{sat},\sqrt{s}\right) = p_{sat}^2 R_{\rm A}^2\ .
\label{eq24}
\end{equation}
In Fig. \ref{fig7} $N_{\rm
AA}\left(b=0,p_0,\sqrt{s}\right)$ is plotted for ${\rm A}=208$ as a function of
$p_0$ at SPS, RHIC and LHC energies. The dashed curve is $p_0^2 R_{\rm
A}^2$. The intersection points give us $p_{sat}$ at the corresponding
energies. Of course, all this is only valid as long as $p_0\gg
\Lambda_{\rm QCD}$ for perturbation theory to be justifiedly used, which
is doubtful at SPS and RHIC (see the Figure; the saturation
momentum are $\sim 0.5$, $\sim 1.4$ and $\sim 2.3$ GeV/c at SPS, RHIC
and LHC energies respectively).

\begin{figure}[htbp]
\vspace{5.3cm}
\hspace{0.5cm}
\epsfysize=12cm
\centerline{\epsffile{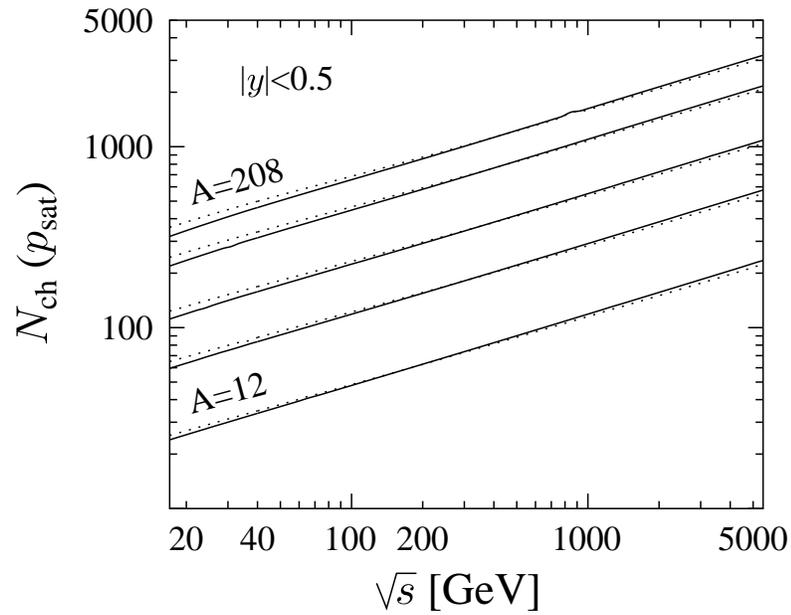}}
\vspace{-8.4cm}
\caption{Rapidity density of charged particles at $y=0$ for ${\rm
A}=12$, 32, 64, 136 and 208 as a function of $\sqrt{s}$ for central A-A
collisions, taken from \protect{\cite{eskola}}. The dotted line is the
result of (\protect{\ref{eq27}}), the solid line is the result of a more
rigorous calculation.}
\label{fig8}
\end{figure}

The values of $N_i=N_{\rm
AA}\left(b=0,p_{sat},\sqrt{s}\right)$ and $p_{sat}$ can be well fitted by the
expressions
\begin{eqnarray}
N_i&=& 1.383\ {\rm A}^{0.922}\ (\sqrt{s})^{0.383}\ ,\label{eq25} \\
p_{sat}&=& 0.208\  {\rm A}^{0.128}\ (\sqrt{s})^{0.191}\ \ {\rm GeV/c}.
\label{eq26}
\end{eqnarray}
The initial state computed in this way very nearly fulfills
the kinetic thermalization
condition for bosons, $\epsilon/n=2.7\  T$ (the number of gluons is much
larger than that of quarks), and there also is some justification to
consider that
further hydrodynamical expansion is locally thermal, i.e. entropy
conserving. Thus initially the entropy $S_i=3.6\  N_i$ (ideal system of
bosons). For the final hadronic gas $S_i=S_f\simeq 4\  N_f$ so that
$N_f=0.9\ N_i$, i.e. the number of hadrons in the final state is, up to
10 \% corrections, equal to the number of initially produced gluons at
the scale $p_{sat}$. The multiplicity prediction \cite{eskola,lastcall}
\begin{equation}
N_{ch}= \frac{2}{3}\ 0.9\ N_i
\label{eq27}
\end{equation}
is directly obtained from (\ref{eq25}) and plotted as a dashed line
in Fig. \ref{fig8}.
The values at RHIC and LHC for central ($b=0$) Pb-Pb collisions are 900 and
3100, not very different from those obtained by DPM, DPMJET and SFM
on very different grounds\footnote{Theoretical models based on a
semiclassical treatment of gluon radiation \cite{mclerran} by partons in the
colliding nuclei give values compatible with these ones, see
\protect{\cite{venu}}.}~.

\section{Other models}
\noindent
In this Section we would like to comments on some other models. The
fact that the rest of the models are joined together, do not mean at all
that they are less important or successful than the mentioned ones. It
is simply the shortage of space which prevents us from a longer study.

\begin{figure}[htbp]
\vspace*{13pt}
\begin{center}
\epsfig{file=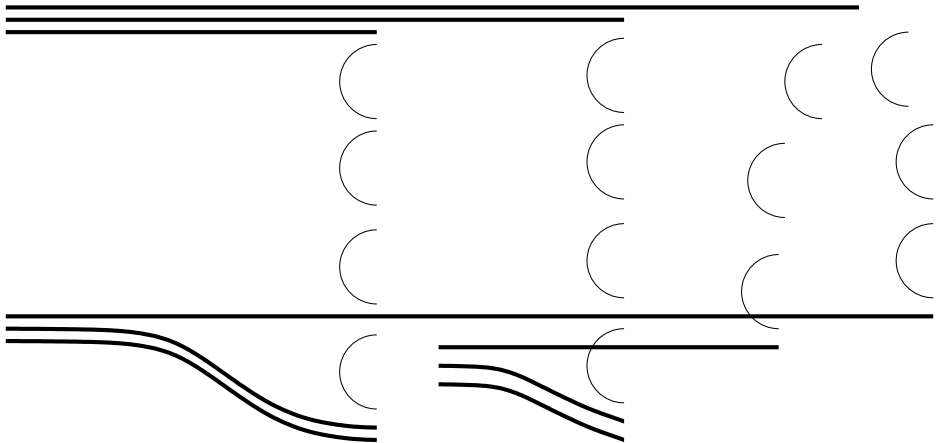,width=12.5cm}
\end{center}
\vspace*{13pt}
\caption{Example of diagram with two color exchanges in VENUS.}
\label{fig9}
\end{figure}

The VENUS model \cite{venus} is an extension of DPM. The main difference
is the inclusion of diagrams in which there is two color exchanges, the
first one providing two $(qq)_v-q_v$ strings, one of the being
intermediate, because a second color exchange breaks the diquark, giving
a different $(qq)_v-q_v$ string
and a double string which consists in a forward
moving quark linked to two backward moving quarks, see Fig. \ref{fig9}.
These diagrams become increasingly important with a growing number of
inelastic collisions, as in h-A or A-B, although of course they are also
present in N-N collisions. They enhance stopping power, shifting the
baryon spectrum towards midrapidities.

VENUS gives large values for central rapidity densities. At $\sqrt{s}=6$
TeV per nucleon for central ($b\leq 3$ fm) Pb-Pb collisions its result is
$dN_{ch}/d\eta|_{\eta=0}=8400$ \cite{alice}.
The model has lately been extended to
deal with $\gamma^*$-$\gamma^*$, $\gamma^*$-h, $\nu$-h, h-h, h-A and A-B
collisions in the same unified approach \cite{hladik}. An unique Pomeron
describes both soft and hard interactions by means of the evolution of
structure functions from some properly chosen initial conditions. The
new model \cite{hladik}, denoted by NEXUS, has not given values for
LHC yet. Preliminary predictions for central ($b<2$ fm) Au-Au collisions at
RHIC are
$dN_{ch}/dy|_{y=0}\simeq 1100$ \cite{lastcall}.
In this model particles and resonances produced in string fragmentation
are allowed to rescatter and, if more than two of them are close enough,
joined into a quark cluster which is decayed isotropically \cite{werner}.

The LUCIAE event generator \cite{luciae} (Lund University and China
Institute of Atomic Energy) is a version of the FRITIOF model
\cite{fritiof} where collective interactions among strings and
rescattering of the produced particles are included. The collective
interactions are incorporated following \cite{and91}. LUCIAE has studied
several observables at SPS comparing to experimental data, but, to our
knowledge, has not worked out predictions for rapidity densities at RHIC
and LHC. In any case, the inclusion of collective string effects
produces very fast particles \cite{andersson}
at the extreme of the phase space, similar
to the cumulative effect which occurs in the SFM. Also, simply by
energy-momentum conservation, a suppression of particles in the central
rapidity region should happen.

\begin{figure}[htbp]
\vspace*{13pt}
\begin{center}
\epsfig{file=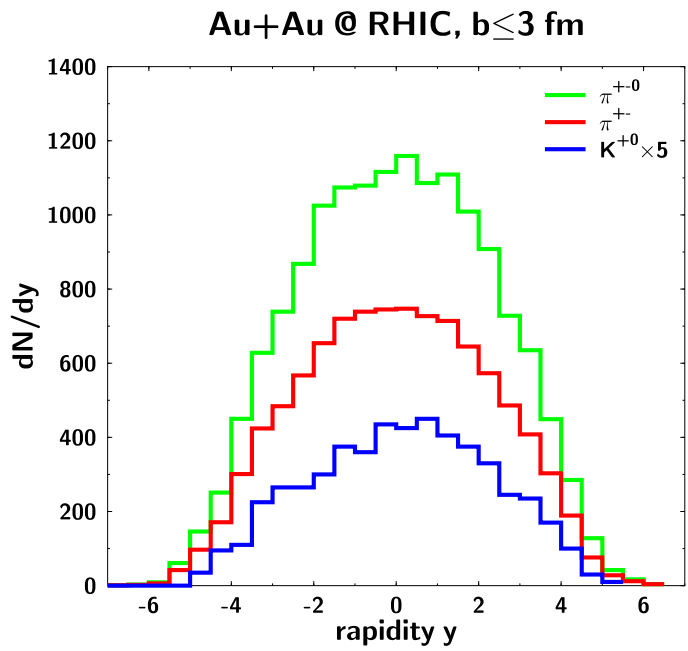,width=11.cm}
\end{center}
\vspace*{13pt}
\vspace*{-0.5cm}
\caption{Predictions for the charged rapidity
density in central ($b\leq 3$ fm)
Au-Au collisions at RHIC in the UrQMD model, taken from
\protect{\cite{urqmd2,lastcall}}.}
\label{fig10}
\end{figure}

The Ultrarelativistic Quantum Molecular Dynamics
(UrQMD) \cite{urqmd1,urqmd2} is a microscopic hadronic approach based on
the covariant propagation of mesonic and baryonic degrees of freedom. It
allows for formation of strings and resonances, and rescattering among
them and of the produced particles. In this aspect it is quite similar
to RQMD. In the low energy region, i.e. $\sqrt{s}\leq 2$ GeV per
nucleon, the inelastic cross sections are dominated by s-channel
formation of resonances, which decay into particles
isotropically in their local rest
frame according to their lifetimes. A large variety of baryonic and
mesonic states have been incorporated in the model. All corresponding
antiparticles are included and treated on the same footing. At higher
energies strings are considered. Much attention is paid in the model to
the intermediate energy region between AGS, the Alternating Gradient
Synchrotron at BNL ($\sqrt{s}\sim 5$ GeV per nucleon),
and SPS, in order to achieve a
smooth transition between the low and high energy regimes. The prediction of
the model \cite{urqmd2,lastcall}
for central ($b\leq 3$ fm) Au-Au collisions at RHIC is shown
in Fig. \ref{fig10}. The central density of charged pions is $\sim 750$,
being for all charged particles $\sim 1100$.

\begin{figure}[htbp]
\vspace*{13pt}
\begin{center}
\epsfig{file=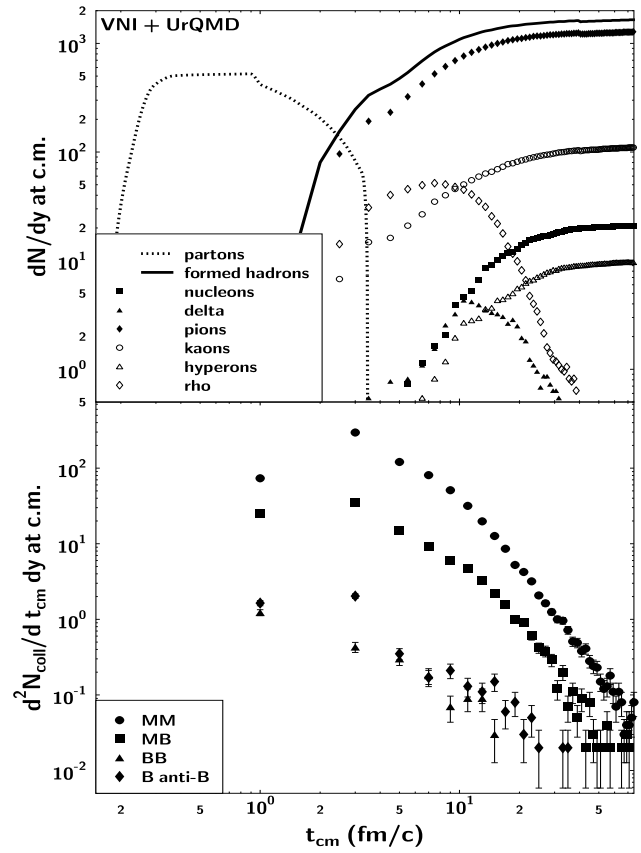,width=9.cm,height=11.5cm}
\end{center}
\vspace*{13pt}
\caption{Time evolution of parton and on-shell hadron rapidity
densities for central
($b\leq 1$ fm) at RHIC  in the VNI+UrQMD model, taken from
\protect{\cite{bass,lastcall}}.}
\label{fig11}
\end{figure}

Another approach using also UrQMD is the model denoted by VNI+UrQMD,
where a combined microscopic partonic/hadronic transport scenario
is introduced
\cite{bass}. The initial high density partonic phase of the heavy ion
reaction is calculated in the framework of the parton cascade model VNI
\cite{vni}, using cross section obtained from PQCD at LO (see
\cite{butvni} for a discussion of the uncertainties introduced by
NLO corrections and
K-factors in VNI and other parton cascade models). The partonic state is
then hadronized via a configuration space coalescence and cluster
hadronization model, and used as initial condition for a hadronic
transport calculation using UrQMD. In Fig. \ref{fig11} the time
evolution of parton and on-shell hadron rapidity densities for central
($b\leq 1$ fm) at RHIC can be seen \cite{bass,lastcall}. From this
curve, the charged particle rapidity density is $\sim 1000$. When,
instead of calculating the initial phase using a parton cascade approach,
QGP formation is assumed, this plasma is evolved hydrodynamically
until hadronization, and then UrQMD is used for hadronic transport
(the model considers a
first order phase transition and is denoted as
Hydro+UrQMD \cite{bass2}), a
smaller charged particle central rapidity density is obtained for
central ($b=0$) Au-Au
collisions at RHIC, $\sim 750$.

\begin{figure}[htbp]
\centerline{\psfig{file=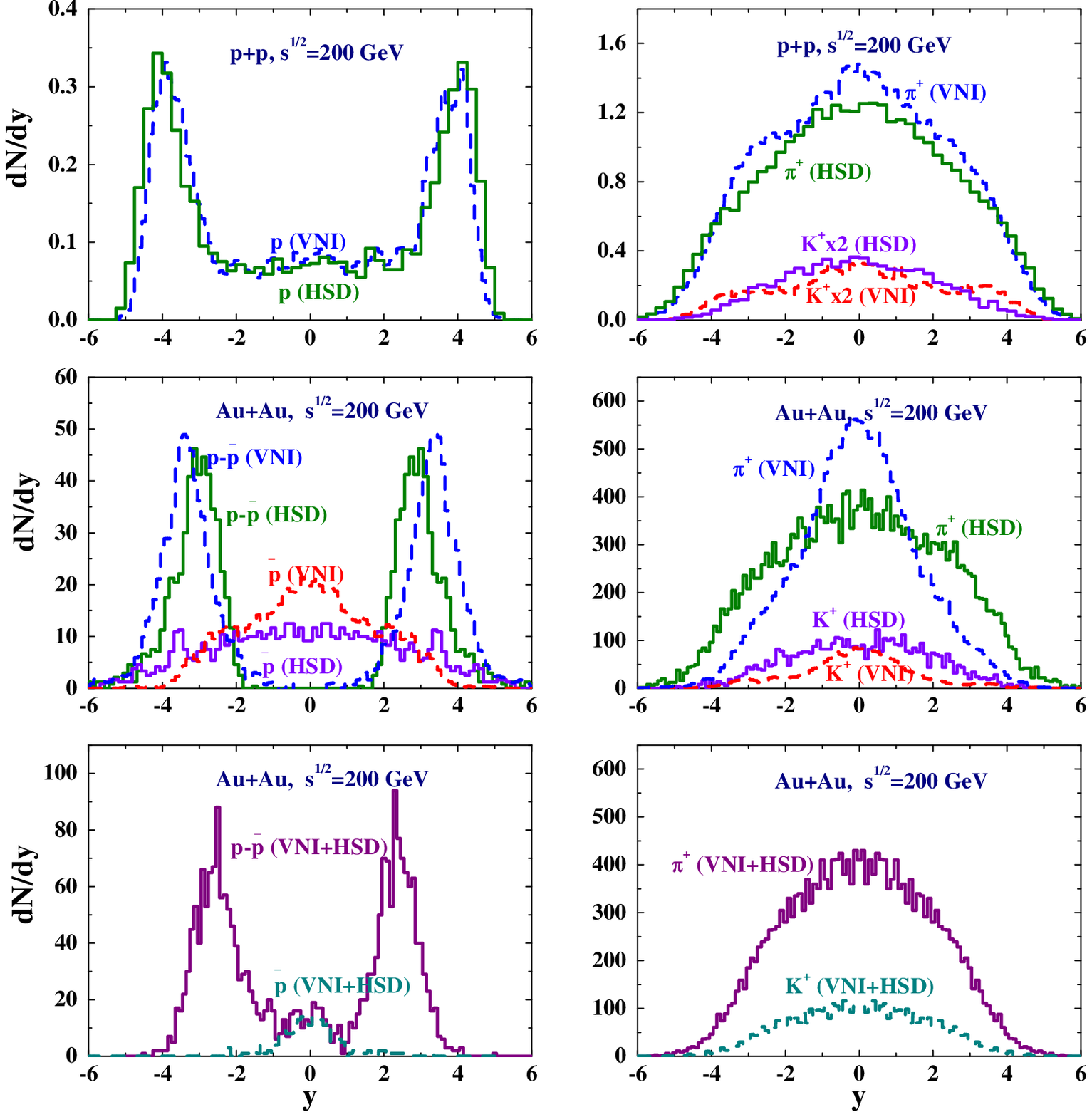,width=16cm}}
\phantom{a}\vspace*{-8.5cm}
\caption{Predictions for rapidity distributions of several particles in
p-p and central ($b\leq 2$ fm) Au-Au collisions at
RHIC from VNI, HSD and VNI+HSD models, taken from \protect{\cite{lastcall}}.}
\label{fig12}
\end{figure}

Using also VNI there is the model denoted by VNI+HSD \cite{cassing},
where HSD stands for Hadron String Dynamics model \cite{cassbra}. This
model involves quarks, diquarks, antiquarks, antidiquarks, strings and
hadrons as degrees of freedom. The parton cascade model VNI is extended
by the hadronic rescattering as described by HSD. Its results  for
central ($b\leq 2$ fm) Au-Au collisions at RHIC are shown
in Fig. \ref{fig12}, where VNI and HSD predictions are the four plots at
the top and
VNI+HSD are the two plots at
the bottom. It is worth noting that VNI+HSD gives almost
the same results as HSD, which already includes final state
interactions. This fact, also observed in DPM and SFM (see Section 2) is
based on unitarity, which controls the number of inelastic collisions
independently of their soft or hard origin. The total number of charged
particles at midrapidity is $\sim 1150$.

Another RHIC prediction \cite{zhang1,lastcall}
comes from a modification of the HIJING model to
include a parton cascade model \cite{zhang2} and final state
interactions based on the ART model \cite{li}. In this model
(HIJING+ZPC+ART) the central rapidity density of charged particles for
central ($b=0$) collisions at RHIC is of the order 1100.

The assumption that local thermodynamical equilibrium is attained by the
system of two heavy ions colliding at high energies is a basic
hypothesis of macroscopic statistical and thermodynamical models
\cite{heinz,becattini,cleymans,bhs} (see \cite{zbsg} for a
discussion on statistical
equilibrium in heavy ion collisions). This idea comes from a long
time ago \cite{fermi,landau,hagedorn}. Following
\cite{brown,bhs,lastcall} the
statistical model treats the system as a grand canonical ensemble with
two free parameters, a temperature $T$ and a chemical potential $\mu_B$.
Interactions of the produced particles are taken into account
considering a excluded volume correction, corresponding to repulsion
setting in for all hadrons at a radius of 0.3 fm (a hard core). Hadron
yield ratios resulting from this model are in reasonable agreement with
SPS central Pb-Pb data. Taking these results and looking at the expected
phase boundary between the QGP and the hadron gas, the hadrochemical
freeze-out points are where one expects, this being suggestive that
hadron yields are frozen at the point when hadronization of the QGP is
complete. This gives for RHIC a freeze-out temperature of 170 MeV, the
same as found at SPS. The chemical potential is expected to be small,
10 MeV used as an upper limit. Strangeness and $I_3$ conservation
then require values of $\mu_s=2.5$ MeV and $\mu_{I_3}=-0.2$ MeV. In
order to predict absolute yields one has to estimate the volume per unit
rapidity at the time when hadronization is complete. Starting from an
initial temperature of $T_i=500$ MeV at a time $\tau=0.2$ fm/c and using
a transverse expansion with $\beta=0.16$, this volume is 3600 fm$^3$ at
the freeze-out temperature of 170 MeV. The number of charged pions per
unit rapidity at $y=0$ is 1260, that of charged kaons 194, that of
protons 62 and that of antiprotons 56. Modifying the freeze-out
temperature to 160 MeV results in a reduction of hadron yields about 10
\%. For central Pb-Pb collisions at the
LHC, taking again $T_f=170$ MeV and $\mu_B=10$ MeV and performing
similar calculations, the fireball volume per unit rapidity at chemical
freeze-out is 14400 fm$^3$, resulting in 5000 charged pions, 770 charged
kaons, 250 protons and 220 antiprotons per unit rapidity. The total
charged particle density is
\begin{equation}
\left.\frac{dN}{dy}\right|_{y=0}=7560.
\label{eq28}
\end{equation}
As at RHIC,
small modifications of $T_f$ and $\mu_B$ lead to small changes in this
prediction. Both at RHIC and LHC the centrality of
these results should correspond roughly to the centrality of the SPS
experimental data which were fitted to extract the parameters used in
the
predictions. As different experiments consider different centrality
criteria,
this is not fully determined, so let us take $5\div 10$ \% as an
estimate.

The quark coalescence model \cite{lastcall} assumes that at RHIC a
QGP will be produced in the collision, which will expand and cool,
hadronization proceeding via quark coalescence as described by the ALCOR
model \cite{blz}. In this nonlinear coalescence model, subprocesses
are not independent, competing one with each other. The coalescence
equations relate the number of hadrons of a given type to the product of
the numbers of different quarks from which the hadron consists. The main
predictions of the model relates different particle ratios, but
unfortunately the absolute values cannot be obtained in the model.

Finally, let us mention an extrapolation done by the WA98 Collaboration
\cite{wa98} from central pseudorapidity densities measured at SPS at different
centralities. The predicted maximum charged pseudorapidity density
for central Pb-Pb collisions is 1000 at RHIC and 2500 at LHC, for $\sim
10$ \%
more central events.

\section{Percolation}
\noindent
In many models, multiparticle production at high energies is described
in terms of color strings stretched between the projectile and target,
see previous Sections. In principle, these strings fragment
independently, the only correlation among them being energy-momentum
conservation. However, with growing energy, centrality
and/or size of the colliding
particles, the number of strings grows and one expects that the
hypothesis of independent fragmentation is no longer valid, interaction
among them becoming essential. For these reasons we have seen in Section
3 and 4 different ways of taking into account such interaction. In
particular, in SFM or RQMD the strings fuse or form color ropes, in such
a way that the transverse size of the new string or rope is the same as
that of the original strings. In this case it can be shown that there is
no phase transition. However, other possibilities could be discussed.
It could be the case that the new string would have a transverse size
corresponding to the sum of the sizes of the original strings. 
In this case, a first order phase transition occurs \cite{bpr}.

An alternative and natural way to the formation of a (non-thermal) QGP
is percolation of strings \cite{abfp,nardi,satzqm99}, which in any case
can be used as an estimation for the failure of the independent
fragmentation hypothesis. This a purely
classical, geometrical mechanism. At a given energy and impact parameter
$b$ in an A-B collision, there is an available transverse area. For
simplicity, let us
take A=B and $b=0$, so this area is $\pi R_{\rm A}^2$. Inside it, the
strings formed in the collision can be viewed as circles of radius $r_0$.
Some of the circles may overlap and then form clusters of strings. Above
a critical density of strings, percolation occurs,
so that clusters of overlapping strings are formed through
the whole collision area. Percolation gives rise to the formation of a
collective state, which can be identified as QGP, at a nuclear scale.
The phenomenon of continuum percolation is well known and has been used
to explain many different physical processes \cite{isi}. The percolation
threshold $\eta_c$ is related to the critical density of strings
$n_c$ by
\begin{eqnarray}
\eta_c&=&\pi r_0^2\ n_c\ ,\label{eq29}\\ 
n_c&=&\frac{N_c}{\pi R_{\rm A}^2}\ \ {\rm at}\ \ b=0,
\label{eq30}
\end{eqnarray}
where $N_c$ is the number of exchanged strings and $r_0\simeq 0.2\div
0.25$ fm \cite{abfp,nardi,shuryak,bali},
see Section 3. $\eta_c$ has been computed using
Monte Carlo simulation and direct connectedness expansions \cite{isi}.
The results lie in the range $1.12\div 1.18$ using step functions for
the profile function of the nucleus (i.e. strings homogeneously
distributed in the whole transverse area available). The use of Woods-Saxon or
Gaussian nuclear densities leads to higher values of $\eta_c$ \cite{dias},
up to $\sim 1.5$. The corresponding value of $n_c$ lies in the range
$6\div 12$
strings/fm$^2$. In Table \ref{tab1} we show the number of strings
exchanged in different central ($b=0$) collisions together with their densities,
as obtained in the SFM \cite{us}. It is seen that percolation could
already occur for Pb-Pb at SPS. At RHIC and LHC, even collisions between
much lighter nuclei could lead to the phase transition.

\begin{table}[htbp]
\caption{Number of strings (upper numbers) and their densities (in
strings/fm$^2$, lower numbers) for different central ($b=0$) collisions
at SPS, RHIC and LHC energies, as obtained in the SFM \protect{\cite{us}}.}
\label{tab1}
\centerline{\small
\begin{tabular}{cccc}\\
\hline\hline
$\sqrt{s}$ (GeV per nucleon) & p-p & S-S & Pb-Pb\\
\hline\hline
19.4 & 3.4 & 144 & 1365 \\
 & 1.7 & 3.4 & 9.2 \\ \hline
200 & 4.3 & 223 & 2029  \\
 & 2.1 & 5.2 & 13.7 \\ \hline
5500 & 5.8 & 416 & 3469 \\
 & 2.9 & 9.8 & 23.4\\
\hline\hline
\end{tabular}}
\end{table}

Notice that string percolation occurs in two dimensions. Percolation of
hadrons was proposed long ago \cite{baym} as a possible way to reach
QGP. However, in this case percolation is three-dimensional and the
critical density is below even normal nuclear matter density. This is in
agreement with lattice studies \cite{bali} which show that the so-called
energy radius of the hadron is about 0.2 fm. Therefore the color fields
inside hadrons occupy only a few percent of the transverse area,
$(r_0/R_h)^2\approx (1/5)^2$. This also explains the relative weak
string-string interaction (for instance the triple Pomeron coupling of
Glauber-Gribov theory used in DPM and DPMJET, see Section 2).

Percolation is a second order phase transition \cite{isi}. The corresponding
scaling law gives the behavior of the number of clusters of $n$ strings,
$\nu_n$, in terms of $\eta$,
\begin{equation}
\langle\nu_n \rangle = n^{-\tau} {\cal F}\left(n^\sigma\left[
\eta- \eta_c\right]\right),\ \ |\eta-
\eta_c|\ll 1, \ \ n\gg 1,
\label{eq31}
\end{equation}
where $\tau=187/91$ and $\sigma=36/91$. The fraction $\phi$ of the total
surface occupied by strings is determined by
\begin{equation}
\phi=1-e^{-\eta}.
\label{eq32}
\end{equation}
It can be seen \cite{bpr} that the multiplicity $\mu_n$ due to a cluster
of $n$ overlapping strings, compared to the multiplicity of one string
$\mu_1$, is given by
\begin{eqnarray}
\mu_n&=&n\ \mu_1\ F(\eta),\label{eq33}\\
F(\eta)&=&\sqrt{\frac{1-e^{-\eta}}{\eta}}\ . \label{eq34}
\end{eqnarray}
From Table \ref{tab1} and taking $r_0=0.25$ fm,
the values of $F(\eta)$ for central ($b=0$) Pb-Pb collisions at RHIC and LHC
are 0.59 and 0.46 respectively, quite close to the values of the
reduction factor of multiplicities in DPM due to triple Pomeron
couplings, 1/2 and 1/3 respectively, see Section 2. A naive calculation
for central rapidity densities of charged particles
can be done multiplying the corresponding
values obtained in the SFM \cite{us} without fusion of
strings for Pb-Pb at  $b=0$, by these reduction factors $F(\eta)$.
The results for RHIC and LHC
are 910 and 1980 respectively (380 at $\sqrt{s}=19.4$ GeV with
$F(\eta)=0.68$) and clearly, from the way they were obtained, should be
considered as lower bounds.

Percolation, in addition to reduce central rapidity multiplicities and enhance
heavy flavor production (as string fusion does),
also gives rise to other consequences in long range
correlations and transverse momentum correlations \cite{abfpp}, and
J/$\psi$ suppression \cite{nardi,dias}. Also, as the energy-momentum of
the clusters is the sum of the energy-momentum of the original strings,
a huge cumulative effect is expected.

The critical point of percolation is the fixed point of a scale
transformation (renormalization group equation) which eliminates short
range correlations, surviving only long range ones. Close the that point
observables should depend only on $\eta$, and not on energy or nuclear
size separately.

By passing, let us notice that a similar intrinsic scale has been
proposed \cite{mclerran} in the small $x$ physics domain, related with
saturation of structure functions or minijets, which was discussed in
Section 6. This quantity is defined by
\begin{equation}
\Lambda^2=\frac{N_{\rm AA}(p_{sat})}{\pi R^2_{\rm A}}\ ,
\label{eq35}
\end{equation}
with $p_{sat}$ the transverse momentum at which saturation starts. When the
number of partons $N_{\rm AA}(p_{sat})$, each one with a transverse size
of the order $\pi/p_{sat}^2$, verifies $\Lambda^2 \pi \simeq p_{sat}^2$, partons
cover the whole nuclear area. Physics should depend only on the value of
$\Lambda^2$. Furthermore, the effective action \cite{mclerran} which
describes small $x$ physics should become critical at some fixed point
of some renormalization group, the correlation functions depending only
on critical exponents determined by symmetry considerations and
dimensionality. Comparing with percolation, indeed $\eta/(\pi r_0^2) =
N/(\pi R_{\rm A}^2)$ is formally $\Lambda^2$, with the exchange of $N$
soft strings by $N_{\rm AA}(p_{sat})$ partons. Let us indicate that from
the arguments
exposed above, overlapping partons will only cover the whole transverse
area $\pi R_{\rm A}^2$ asymptotically. According to (\ref{eq32}), the
transverse area $S$
covered by $N_{\rm AA}(p_{sat})$ partons of transverse size $\pi/p_{sat}^2$
is
\begin{equation}
S=\pi R_{\rm A}^2\ \left[1-\exp{\left(-\Lambda^2\pi/r_0^2\right)}\right].
\label{eq36}
\end{equation}

\section{Cosmic Ray Physics and heavy ion accelerators}
\noindent
Usually it is considered that the highest cosmic ray energies, say
$10^{15}\div 10^{20}$ eV, are much higher than energies reached at
accelerators. With the advent of RHIC and LHC, this is not true any
longer. Pb-Pb collisions at RHIC and LHC will reach total energies of
$\sim 10^{15}$ and $\sim 10^{18}$ eV respectively. This means that, although no
participant nuclei larger than Fe is expected, there will be collective
physics to explore in Cosmic Ray Physics.
In particular changes in the multiplicity originate
\cite{gaisser} changes in the development of atmospheric showers.
Unfortunately, atmospheric showers are dominated by forward particles,
and it is in the fragmentation regions where models which in the
central rapidity region are quite different, are more similar.
Nevertheless, collective effects like color rope formation will
influence the fragmentation regions (for example, the enhancement of
the
cumulative effect) and have
observable effects in the development of the shower
\cite{cosmic1,cosmic2}. Besides, PQCD effects may be of importance for the
transverse broadening of the shower. It would be convenient to apply
the different models for multiparticle production to simulations of
cosmic ray atmospheric showers \cite{cosmic3,cosmic4}.

\section{Conclusions}
\noindent
In Table \ref{tab2} predictions of the different models for charged
particle densities produced in central Au-Au or Pb-Pb collisions at RHIC
and LHC are presented, together with the corresponding centrality criteria.
Some of the
predictions of the models are not available and its place has been left
empty.
In order to estimate the discrepancy between different
predictions induced by different definitions of centrality,
by the fact that
some of the results are $dN/dy|_{y=0}$ and other $dN/d\eta|_{\eta=0}$,
and by using Au-Au or Pb-Pb at RHIC or slightly different energies at
LHC, in Table
\ref{tab3} we present results obtained in the SFM \cite{us} for
different reactions. From them it can be concluded that results should
be compared allowing for a $20\div 30$ \% discrepancy.

For completeness, we have included predictions from percolation
and from the WNM \cite{bialas}. As discussed in Section 8, the former should be
considered as lower bounds. The latter have been obtained
computing in SFM \cite{us} $dN_{ch}/dy|_{y=0}$ for p-p, p-n and n-n
collisions at the corresponding energies, making an isospin weighted
average and
multiplying by the number of wounded nucleons in a central collision,
taking $\overline{n}_{\rm A}=200$ in Pb-Pb and $\overline{n}_{\rm
A}=190$ in Au-Au (which correspond to $b\simeq 0$); the result that we
get 
in this way for Pb-Pb at $\sqrt{s}=17.3$ GeV per nucleon (SPS) is 370.
In the following we will not discuss these two quite naive predictions.

\begin{table}[htbp]
\caption{Predictions of different model for charged particle densities in
central collisions. Unless otherwise stated, predictions refer to
$dN/dy$ at $y=0$, for Au-Au
at $\sqrt{s}=200$ GeV per nucleon at RHIC and Pb-Pb at $\sqrt{s}=5.5$
TeV per nucleon at LHC. The centrality criteria and the references from
which the results have been taken are indicated in each
case.}
\label{tab2}
\centerline{\small
\begin{tabular}{ccc}\\
\hline\hline
Model & RHIC & LHC\\
\hline\hline
DPM
 & 1000 & 2500 \\
($\overline{n}_{\rm A}>200$) \protect{\cite{ckat}} & (Pb-Pb) &
($\sqrt{s}=7$ TeV per nucleon)\\ \hline
DPMJET
 & 1280 & 2800 \\
\protect{\cite{dpmjet2}} & (Pb-Pb, 3 \%) & (4 \%) \\ \hline
SFM
 & 910 & 3140  \\
(5 \%) \protect{\cite{us}} & & \\ \hline
RQMD
 & 700 &  \\
\protect{\cite{lastcall,sorgeqm99}} & ($b=3$ fm) & \\ \hline
HIJING
 & $600\div 1150$ & $5000\div 7500$ \\
($dN/d\eta$ at $\eta=0$, $b<3$ fm)
\protect{\cite{lastcall,hijing}}& &\\ \hline
Eskola {\it et al.}
 & 900  & 3100 \\
($b=0$) \protect{\cite{eskola}} & (Pb-Pb) & \\ \hline
HIJING+ZPC+ART
 & 1100 &  \\
\protect{\cite{zhang1,lastcall}} & ($b=0$) & \\ \hline
UrQMD
 & 1100 & \\
\protect{\cite{urqmd2,lastcall}} &  ($b\leq 3$ fm) & \\ \hline
VNI+UrQMD
 & 1000 &  \\
\protect{\cite{bass,lastcall}} & ($b\leq 1$ fm) & \\ \hline
Hydro+UrQMD
 & 750 &  \\
\protect{\cite{bass2}} & ($b=0$) & \\ \hline
VNI+HSD & 1150 & \\
\protect{\cite{lastcall}} & ($b\leq 2$ fm) & \\ \hline
VENUS 4.12
 & & 8400 ($dN/d\eta$ at $\eta=0$, \\
\protect{\cite{alice}} & & $b\leq 3$ fm, $\sqrt{s}=6$ TeV per nucleon) \\
\hline
NEXUS &
1100 & \\
\protect{\cite{lastcall}} & ($b<2$ fm) & \\ \hline 
Statistical & 1550 & 7560 \\
($5\div 10$ \%) \cite{brown,bhs,lastcall} & & \\ \hline
WA98 extrapolation & 1000 & 2500 \\
(maximum $dN/d\eta$, $\sim 10$ \%) \cite{wa98} & (Pb-Pb) & \\ \hline
WNM & 560 & 1220 \\
 & ($\overline{n}_{\rm A}=190$) & ($\overline{n}_{\rm A}=200$) \\ \hline
Percolation  & 910 & 1980 \\
($b=0$) & (Pb-Pb) & \\ \hline\hline
\end{tabular}}
\end{table}

\begin{table}[htbp]
\caption{Results obtained in the SFM \protect{\cite{us}}
for charged particle production in different reactions,
compared with the results for $dN/dy$ at $y=0$ in Au-Au at
$\sqrt{s}=200$ GeV per nucleon (910) and in Pb-Pb at 
$\sqrt{s}=5.5$ TeV per nucleon (3140), with a degree of centrality of 5
\%.}
\label{tab3}
\centerline{\small
\begin{tabular}{ccc}\\
\hline\hline
Reaction and observable & Result & Difference (\%)\\
\hline\hline
$dN/dy$ at $y=0$, Au-Au at $\sqrt{s}=200$ GeV & 910 & $0$ \\
per nucleon, 5 \% of centrality ($b\leq 3.2$ fm) & &\\
\hline
$dN/d\eta$ at $\eta=0$, Au-Au at $\sqrt{s}=200$ GeV & 730 & $-20$ \\
per nucleon, 5 \% of centrality ($b\leq 3.2$ fm) & &\\
\hline
Maximum $dN/d\eta$, Au-Au at $\sqrt{s}=200$ GeV & 770 & $-15$ \\
per nucleon, 5 \% of centrality ($b\leq 3.2$ fm) & &\\
\hline
$dN/dy$ at $y=0$, Au-Au at $\sqrt{s}=200$ GeV & 1030 & $+13$ \\
per nucleon, $b=0$ & &\\
\hline
$dN/dy$ at $y=0$, Pb-Pb at $\sqrt{s}=200$ GeV & 960 & $+5$ \\
per nucleon, $b\leq 3.2$ fm & &\\
\hline
$dN/dy$ at $y=0$, Pb-Pb at $\sqrt{s}=5.5$ TeV & 3140 & $0$ \\
per nucleon, 5 \% of centrality ($b\leq 3.3$ fm) & &\\
\hline
$dN/d\eta$ at $\eta=0$, Pb-Pb at $\sqrt{s}=5.5$ TeV & 2620 & $-17$ \\
per nucleon, 5 \% of centrality ($b\leq 3.3$ fm) & &\\
\hline
Maximum $dN/d\eta$, Pb-Pb at $\sqrt{s}=5.5$ TeV & 2820 & $-10$ \\
per nucleon, 5 \% of centrality ($b\leq 3.3$ fm) & &\\
\hline
$dN/dy$ at $y=0$, Pb-Pb at $\sqrt{s}=5.5$ TeV & 3590 & $+14$ \\
per nucleon, $b= 0$ & &\\
\hline
$dN/dy$ at $y=0$, Pb-Pb at $\sqrt{s}=7$ TeV & 3330 & $+6$ \\
per nucleon, 5 \% of centrality ($b\leq 3.3$ fm) & &\\
\hline\hline
\end{tabular}}
\end{table}

At RHIC energies all the predictions are in the range $700\div 1550$ and
most of them in $1000\div 1100$. The lowest value corresponds to RQMD.
The reason for that, as already commented at the end of Section 4, is the
formation of color ropes with a large probability, which is required in
the model to describe antibaryon enhancement at SPS because
a rather large baryon-antibaryon annihilation cross section is
used. In SFM, which considers a similar mechanism (string fusion), the
charged density
obtained is larger. The low value also obtained in Hydro+UrQMD is due
to the initial QGP state and the first order transition assumed.

At LHC the differences among the predictions are larger, more than a
factor 3. They lie in the
range $2500\div 8500$. Comparing Table \ref{tab2} with Fig. \ref{fig1},
which summarizes the situation before 1996, it can be seen that nowadays
predictions tend to gather around the lowest values; at the time of
\cite{alice} only SFM gave predictions below 4000.

Essentially, models based on parton shower evolution predict larger
values. Also statistical models obtain very large charged densities.
However, the model described in Section 6, which uses mainly partonic
degrees of freedom, obtains a low value (3100). This is due to the
saturation of minijet production, which plays the role of an upper
cut-off in the number of minijets. This saturation is a consequence of
unitarity \cite{mueller}.

Unitarity is a basic ingredient that controls the number of soft
and hard elementary scatterings (which are no more
independent one from each other)
in models like DPM, DPMJET, SFM and
RQMD. In addition to that, energy-momentum conservation reduces the
possible number of scatterings. Finally,
interaction among strings is another collective
effect which reduces central pion
densities. In DPM these interactions are taken into
account by means of the triple Pomeron. Its coupling is fixed to
describe soft diffraction and HERA data. In other models as RQMD or SFM
the interactions among strings are taken into account via the formation
of color ropes or fusion of strings. Its strength is essentially fixed
to reproduce heavy flavor and antibaryon enhancement at SPS. It is not
unexpected that these three different
forms of quantifying the shadowing give rise
to similar predictions for global and simple observables as central
rapidity densities of charged particles. Probably the knowledge of
shadowing from small $x$ physics can help to reconcile models based on
partonic degrees of freedom with those based on strings as degrees
of freedom.
On the other hand, statistical thermal models predict a larger value
close to 8000 as a consequence of the rather large volume at freeze-out.
A reduction in a factor 3 would mean a strong reduction in this volume
and a large change of the ratio between different particles, or in the
temperature and chemical potential values.

In the summary talks of three mayor conferences
\cite{qm992,brown,eskolacern} were quoted 8000, 8000, and between 3000
and 8000, for the probable number of charged particles per unit rapidity
at the center of mass in central Pb-Pb collisions at the LHC. Models
with interaction among strings obtain a lower value. The interplay
between soft and hard physics is one of the main issues in the study of
strong interactions and, together with the search and characterization of
QGP, one of the main goals of Heavy Ion Physics. Doubtless, the new
experiments at RHIC and LHC will shed light on this subject, even
measuring such a simple observable as central rapidity
densities of charged particles.

\section*{Acknowledgements}
\noindent
This work has been done under contract AEN99-0589-C02 of CICYT (Spain).
We express
our gratitude to N. S. Amelin, M. A. Braun, A. Capella, W. Cassing, K.
J. Eskola, E. G. Ferreiro, A. B. Kaidalov, J. Ranft, C. A. Salgado, H.
Sorge, D. Sousa and K. Werner for useful discussions and comments.

\newpage

\centerline{\Large \bf FIGURES}

\begin{thebibliography}{000}

\bibitem{hwa} {\it Quark--Gluon Plasma}, ed. R. C. Hwa
(World Scientific, Singapore,
1990); {\it Quark--Gluon Plasma 2}, ed. R. C. Hwa 
(World Scientific, Singapore,
1995).

\bibitem{books} C.-Y. Wong, {\it Introduction to High-Energy
Heavy-Ion Collisions}
(World Scientific, Singapore, 1994); L. P. Csernai, {\it Introduction to
Relativistic Heavy-Ion Collisions}
(John Wiley \& Sons, Chichester, 1994).

\bibitem{qm991} J.-P. Blaizot, in {\it Proceedings of Quark Matter 99}
(Torino, Italy, May 10th-15th 1999), to appear in {\it Nucl. Phys.}
A (hep-ph/9909434, 1999).

\bibitem{qm992} P. Braun-Munzinger, in {\it Proceedings of Quark
Matter 99}
(Torino, Italy, May 10th-15th 1999), to appear in {\it Nucl. Phys.}
A (nucl-ex/9908007,
1999).

\bibitem{brown} J. Stachel, in {\it Proceedings of the XXIX
International Symposium on Multiparticle Dynamics}
(Providence, USA,
August 9th-13th 1999), to be published by World Scientific.

\bibitem{c1} C. Pajares, {\it Acta Physica Polonica} B30 (1999)
2263.

\bibitem{ms} T. Matsui and H. Satz, {\it Phys. Lett.} B178 (1986)
416.

\bibitem{rm} J. Rafelski and B. M\"uller,
{\it Phys. Rev. Lett.} 48 (1982) 1066; Erratum: {\it ibid.} 56
(1986) 2334.

\bibitem{raf} J. Rafelski, {\it
Phys. Rep.} 88 (1982) 331.

\bibitem{na50} NA50 Collaboration: M. C. Abreu {\it et al.}, {\it
Phys. Lett.} B410 (1997) 327; {\it ibid.} 337.

\bibitem{strange} WA97 Collaboration: E. Andersen {\it et al.},
{\it Phys. Lett.} B433 (1998) 209;
NA49 Collaboration: H. Appelshauser {\it et al.},
{\it ibid.} B444 (1998) 523; WA85 Collaboration: F.
Antinori {\it et al.},
{\it ibid.} B447 (1999) 178.

\bibitem{dilep} CERES Collaboration: G. Agakichiev {\it et al.}, 
{\it Phys. Rev. Lett.} 75 (1995) 1272; {\it Phys. Lett.} B422 (1998) 405.

\bibitem{bo} J.-P. Blaizot and J.-Y. Ollitrault, {\it Phys. Rev.
Lett.} 77 (1996) 1703.

\bibitem{wong} C.-Y. Wong, {\it Phys. Rev.} C55 (1997) 2621.

\bibitem{kns} D. Kharzeev, M. Nardi and H. Satz, {\it Z. Phys.} C74
(1997) 307.

\bibitem{abfp} N. Armesto, M. A. Braun, E. G. Ferreiro and C. Pajares,
{\it Phys. Rev. Lett.} 77 (1996) 3736.

\bibitem{acf} N. Armesto, A. Capella and E. G. Ferreiro, {\it Phys.
Rev.} C59 (1999) 395.

\bibitem{ggbcm} J. Geiss, C. Greiner, E. L. Bratkovskaya, W. Cassing and U.
Mosel,
{\it Phys. Lett.} B447 (1999) 31.

\bibitem{vgw}
S. E. Vance, M. Gyulassy and X.-N. Wang,
{\it Phys. Lett.} B443 (1998) 45.

\bibitem{cfs} A. Capella, E. G. Ferreiro and C. A. Salgado, {\it
Phys. Lett.} B459 (1999) 27.

\bibitem{lkb} G. Q. Li, C. M. Ko and G. E. Brown,
{\it Phys. Rev. Lett.} 75 (1995) 4007.

\bibitem{alice} ALICE Collaboration: {\it
Technical Proposal for A Large Ion Collider
Experiment at the CERN LHC}, preprint CERN/LHCC/95-71
(1995). 

\bibitem{sfm} N. S. Amelin, M. A. Braun and C. Pajares,
{\it Phys. Lett.} B306 (1993) 312;
{\it Z. Phys.} C63 (1994) 507. 

\bibitem{venus} K. Werner, {\it Phys. Rep.} 232 (1993) 87. 

\bibitem{shaker} N. van Eijndhoven {\it et al.}, Internal Note ALICE
95-32 (1995).

\bibitem{lastcall} S. A. Bass {\it et al.}, {\it Last Call for
RHIC Predictions}, in
{\it
Proceedings of Quark Matter 99}
(Torino, Italy, May 10th-15th 1999), to appear in {\it Nucl. Phys.}
A (nucl-th/9907090, 1999).

\bibitem{dpm} A. Capella, U. P. Sukhatme, C.-I. Tan and J. Tran Thanh
Van, {\it Phys. Lett.} 81B (1979) 69; {\it Phys. Rep.} 236 (1994)
225;  A. Capella, preprint LPT-Orsay 99-75 (hep-ph/9910219, 1999).

\bibitem{ckt} A. Capella, J. Kwiecinski and J. Tran Thanh Van, {\it
Phys. Lett.} B108 (1982) 347. 

\bibitem{cpr} A. Capella, C. Pajares and A. V. Ramallo, {\it Nucl.
Phys.} B241 (1984) 75.

\bibitem{dpmjet1} J. Ranft, {\it Phys. Rev.} D51 (1995) 64.

\bibitem{dpmjet2} J. Ranft, preprint
SI-99-5 (hep-ph/9911213, 1999).

\bibitem{qgsm} A. B. Kaidalov and K. A. Ter-Martirosyan, {\it Phys.
Lett.} B117 (1982) 247.

\bibitem{fritiof} B. Andersson, G. Gustafson and B. Nilsson-Almqvist,
{\it Nucl. Phys.} B281 (1987) 289; B. Nilsson-Almqvist and
E. Stenlund,
{\it Comput. Phys. Commun.} 43 (1987) 387.

\bibitem{abctg} N. S. Amelin, L. V. Bravina, L. P. Csernai,
V. D. Toneev, K. K. Gudima and  S. Yu.
Sivoklokov,
{\it Phys. Rev.} C47 (1993) 2299.

\bibitem{us} N. S. Amelin, N. Armesto, C. Pajares and D. Sousa, in
preparation.

\bibitem{rqmd1} H. Sorge, H. St\"ocker and W. Greiner,
{\it Ann. Phys.} 192 (1989) 266.

\bibitem{rqmd2} H. Sorge, M. Berenguer,
H. St\"ocker and W.
Greiner, {\it Phys. Lett.} B289 (1992) 6.

\bibitem{rqmd3} H. Sorge, {\it Phys.
Rev.}
C52 (1995) 3291.

\bibitem{urqmd1} M. Bleicher {\it et al.},
{\it J. Phys.} G25 (1999) 1859.

\bibitem{urqmd2} M. Bleicher, S. A. Bass, H. St\"ocker
and W.
Greiner, in {\it Proceedings of Quark Matter 99}
(Torino, Italy, May 10th-15th 1999), to appear in {\it Nucl. Phys.}
A (hep-ph/9906398, 1999).

\bibitem{hladik} M. Hladik, PhD. Thesis, Nantes, December 1999;
H. J. Drescher, M. Hladik, S. Ostapchenko and
K. Werner, {\it J. Phys.} G25 (1999) L91;
in
{\it
Proceedings of Quark Matter 99}
(Torino, Italy, May 10th-15th 1999), to appear in {\it Nucl. Phys.}
A
(hep-ph/9906428, 1999).

\bibitem{luciae} 
A. Tai, preprint LU-TP-94-22; preprint LU-TP-95-2;
B.-H. Sa, Z.-Q. Wang, X.-Z. Zhang, G. Song, Z.-D. Lu and
Y.-M. Zheng,
{\it Phys. Rev.} C48 (1993) 2995;
B.-H. Sa,
A. Tai and
Z.-D. Lu,
{\it Phys. Rev.} C52 (1995) 2069;
B.-H. Sa and
A. Tai,
{\it Phys. Rev.} C55 (1997) 2010;
C57 (1998) 261.

\bibitem{hijing}
X.-N. Wang and M. Gyulassy,
{\it Phys. Rev.} D44 (1991) 3501;
D45 (1992) 844;
M. Gyulassy and X.-N. Wang,
{\it Comput. Phys. Commun.} 83 (1994) 307; X.-N. Wang, in {\it
Proceedings of Quark Matter 99}
(Torino, Italy, May 10th-15th 1999), to appear in {\it Nucl. Phys.}
A
(nucl-th/9907093, 1999).

\bibitem{eskola} K. J. Eskola and K. Kajantie,
{\it Z. Phys.} C75 (1997) 515;
K. J. Eskola,
{\it Comments Nucl. Part. Phys.} 22 (1998) 185;
K. J. Eskola, K. Kajantie, P. V. Ruuskanen
and K. Tuominen, preprint
JYFL-8-99
(hep-ph/9909456, 1999);
K. J. Eskola and K. Tuominen, in {\it 
Proceedings of Quark Matter 99}
(Torino, Italy, May 10th-15th 1999), to appear in {\it Nucl. Phys.}
A
(hep-ph/9906438, 1999).

\bibitem{geiger}
K. Geiger and B. M\"uller,
{\it Nucl. Phys.} B369 (1992) 600;
K. Geiger,
{\it Phys. Rev.} D47 (1993) 133.

\bibitem{hardinit} J.-P. Blaizot and A. H. Mueller,
{\it Nucl. Phys.} B289 (1987) 847; K. Kajantie, P. V.
Landshoff and J. Lindfors,
{\it Phys. Rev. Lett.} 59 (1987) 2527.

\bibitem{kajan} K. Kajantie, in {\it Proceedings of the
15th International Conference on Particle and
Nuclei (PANIC 99)} (Uppsala, Sweden, June 10th-16th  1999),
hep-ph/9907544 (1999).

\bibitem{mueller}
A. H. Mueller, {\it Nucl. Phys.} A654 (1999) 37;
preprint CU-TP-937 (hep-ph/9904404, 1999); preprint CU-TP-954
(hep-ph/9911289, 1999).

\bibitem{gabriele} G. 't Hooft,
{\it Nucl. Phys.} B72 (1974) 461; G. Veneziano,
{\it Nucl. Phys.} B74 (1974) 365.

\bibitem{reggeon} V. N. Gribov,
{\it Sov. Phys. JETP} 26 (1968) 414.

\bibitem{glauber} R. J. Glauber, in {\it Lectures in Theoretical
Physics}, Vol. 1, ed. W. E. Brittin and L. G. Duham
(Interscience, New York, 1959).

\bibitem{gribov} V. N. Gribov, {\it Sov. Phys. JETP} 29 (1969) 483;
{\it ibid.} 30 (1970) 709.

\bibitem{bbz} A. Bialas, M. Bleszy\'nski and
W. Czyz,
{\it Nucl. Phys.} B111 (1976) 461.

\bibitem{bialas} A. Bialas, in {\it Proceedings of the XIIIth International
Symposium
on Multiparticle
Dynamics}, ed. W. Kittel, W. Metzger and A. Stergiou (World
Scientific,
Singapore, 1983).

\bibitem{na49qm99} NA49 Collaboration: F. Sikl\'er {\it et al.}, in
{\it
Proceedings of Quark Matter 99}
(Torino, Italy, May 10th-15th 1999), to appear in {\it Nucl. Phys.}
A; NA35 Collaboration: T. Alber {\it et al.}, preprint
IKF-HENPG/6-94 (1997); M.
Ga\'zdzicki and D. R\"ohrich, {\it Z. Phys.} C65 (1995) 215.

\bibitem{wa98} WA98 Collaboration: T. Peitzmann {\it et al.},  in
{\it
Proceedings of Quark Matter 99}
(Torino, Italy, May 10th-15th 1999), to appear in {\it Nucl. Phys.}
A.

\bibitem{cmt} A. Capella, C. Merino and J. Tran Thanh Van, {\it Phys.
Lett.} B265 (1991) 415.

\bibitem{ckat} A. Capella, A. B. Kaidalov and J. Tran Thanh Van, {\it
Heavy Ion Physics} 9 (1999) 169.

\bibitem{ckmpt} A. Capella, A. B. Kaidalov, C. Merino, D. Pertermann and
J. Tran Thanh Van, {\it Phys. Rev.} D53 (1996) 2309;
{\it Eur. Phys. J.} C5 (1998) 111.

\bibitem{hard} A. Capella, J.
Kwiecinski and J. Tran Thanh Van,
{\it Phys. Rev. Lett.} 58 (1987) 2015.

\bibitem{zjil} T. Sj\"ostrand and M. van Zijl,
{\it Phys. Rev.} D36 (1987) 2019.

\bibitem{pythia}
H.-U. Bengtsson and T. Sj\"ostrand,
{\it Comput. Phys. Commun.} 46 (1987) 43.

\bibitem{grv} M. Gl\"uck, E. Reya and A. Vogt, {\it
Eur. Phys. J.} C5 (1998) 461.

\bibitem{ck} A. Capella and B. Z. Kopeliovich, {\it Phys. Lett.} B381
(1996) 325;
A. Capella and C. A. Salgado,
{\it Phys. Rev.} C60 (1999) 054906.

\bibitem{vg}
S. E. Vance and M. Gyulassy,
{\it Phys. Rev. Lett.} 83 (1999) 1735; D. Kharzeev, {\it Phys.
Lett.} B378 (1996) 238; B. Z. Kopeliovich and B. Povh, {\it Phys. 
Lett.} B446 (1999) 321; J. A. Casado, {\it Nucl. Phys.} A651 (1999)
93.

\bibitem{rv} G. C. Rossi and G. Veneziano, {\it Nucl. Phys.} B123
(1977) 507.

\bibitem{gross} D. J. Gross and H. Ooguri,
{\it Phys. Rev.} D58 (1998) 106002.

\bibitem{bp} M. A. Braun and C. Pajares,
{\it Phys. Lett.} B287 (1992) 154; {\it
Nucl. Phys.} B390 (1993) 542; {\it ibid.} 559.

\bibitem{ranftmerino} C. Merino, C. Pajares and
J. Ranft,
{\it Phys. Lett.} B276 (1992) 168;
H. J. M\"ohring, J. Ranft, C. Merino and C. Pajares,
{\it Phys. Rev.} D47 (1993) 4142.

\bibitem{schwinger} J. Schwinger,
{\it Phys. Rev.} 82 (1951) 664.

\bibitem{casher} A. Casher, H. Neuberger and S. Nussinov,
{\it Phys. Rev.} D20 (1979) 179.

\bibitem{biro} T. S. Bir\'o, H. B. Nielsen and J. Knoll,
{\it Nucl. Phys.} B245 (1984) 449.

\bibitem{andersson} B. Andersson and  P. A. Henning,
{\it Nucl. Phys.} B355 (1991) 82.

\bibitem{nardi} M. Nardi and H. Satz,
{\it Phys. Lett.} B442 (1998) 14; H. Satz,
{\it Nucl. Phys.} A642 (1998) 130.

\bibitem{bpr} M. A. Braun, C. Pajares and J.
Ranft,
{\it Int. J. Mod. Phys.} A14 (1999) 2689.

\bibitem{abfpp} N. Armesto, M. A. Braun, E. G. Ferreiro and C. Pajares,
in
{\it
Proceedings of Quark Matter 99}
(Torino, Italy, May 10th-15th 1999), to appear in {\it Nucl. Phys.}
A;
M. A. Braun and C. Pajares, preprint
US-FT-15-99
(hep-ph/9907332, 1999), to appear in {\it Eur. Phys. J.} C.

\bibitem{cosmic1} N. Armesto, M. A. Braun, E. G. Ferreiro, C. Pajares
and
Yu. M. Shabelski, {\it Phys. Lett.} B389 (1996) 78;
{\it Astropart. Phys.} 6 (1997) 327.

\bibitem{antilam} N. Armesto, M. A. Braun, E. G. Ferreiro and C.
Pajares,
{\it Phys. Lett.} B344 (1995) 301.

\bibitem{shuryak} E. V. Shuryak, hep-ph/9911244 (1999);
{\it Phys. Lett.} 79B (1978) 135.

\bibitem{bali} G. S. Bali, preprint HUB-EP-98-57
(hep-ph/9809351, 1998);
F. V. Gubarev, E. M. Ilgenfritz, M .I.
Polikarpov and T. Suzuki, preprint
ITEP-TH-43-99
(hep-lat/9909099, 1999).

\bibitem{sjostrand} B. Andersson, G. Gustafson, G. Ingelman and T.
Sj\"ostrand,
{\it Phys. Rep.} 97 (1983) 31;
T. Sj\"ostrand,
{\it Comput. Phys. Commun.} 39 (1986) 347.

\bibitem{sorgeqm99} H. Sorge, in
{\it
Proceedings of Quark Matter 99}
(Torino, Italy, May 10th-15th 1999), to appear in {\it Nucl. Phys.}
A (nucl-th/9906051, 1999).

\bibitem{quenching} X.-N. Wang,
{\it Phys. Rev.} C58 (1998) 2321; X.-N. Wang and M.
Gyulassy,
{\it Phys. Rev. Lett.} 68 (1992) 1480.

\bibitem{eks98} K. J. Eskola, V. J. Kolhinen and C. A. Salgado,
{\it Eur. Phys. J.} C9 (1999) 61.

\bibitem{mclerran} L. McLerran and R. Venugopalan,
{\it Phys. Rev.} D49 (1994) 2233; {\it ibid.} 3352; {\it
ibid.} D50 (1994) 2225;
L. McLerran,
preprint TPI-MINN-99-14
(hep-ph/9903536, 1999).

\bibitem{venu} A. Krasnitz and R. Venugopalan, in {\it Proceedings of
the XXIX
International Symposium on Multiparticle Dynamics}
(Providence, USA,
August 9th-13th 1999), to be published by World Scientific
(hep-ph/9910391, 1999);  hep-ph/9909203 (1999).

\bibitem{werner} K. Werner and J. Aichelin,
{\it Phys. Rev. Lett.} 76 (1996) 1027.

\bibitem{and91} B. Andersson,
{\it Phys. Lett.} B256 (1991) 337.

\bibitem{bass}
S. A. Bass, M. Hofmann, M. Bleicher, L. V. Bravina, E. E. Zabrodin, H.
St\"ocker and W. Greiner,
{\it Phys. Rev.} C60 (1999) 021901.

\bibitem{vni} K. Geiger,
{\it Comput. Phys. Commun.} 104 (1997) 70.

\bibitem{butvni} S. A. Bass and B. M\"uller, preprint DUKE-TH-99-192
(nucl-th/9908014, 1999), to appear in {\it Phys. Lett.} B.

\bibitem{bass2} S. A. Bass, A. Dumitru, M. Bleicher, L. V. Bravina, E.
E. Zabrodin, H. St\"ocker and W. Greiner,
{\it Phys. Rev.} C60 (1999) 021902.

\bibitem{cassing} W. Cassing,
E. L. Bratkovskaya, J. Geiss, C. Greiner, S. Juchem
and U. Mosel,
 in
{\it
Proceedings of Quark Matter 99}
(Torino, Italy, May 10th-15th 1999), to appear in {\it Nucl. Phys.}
A (nucl-th/9906072, 1999).

\bibitem{cassbra} W. Cassing and E. L. Bratkovskaya,
{\it Phys. Rep.} 308 (1999) 65; {\it Nucl. Phys.} A623 (1997)
570.

\bibitem{zhang1}
B. Zhang, C. M. Ko, B.-A. Li and Z. Lin,
nucl-th/9904075 (1999).

\bibitem{zhang2} B. Zhang, {\it Comput. Phys. Commun.} 109 (1998)
193.

\bibitem{li} B.-A. Li and C. M. Ko, {\it Phys. Rev.} C52 (1995) 2037.

\bibitem{heinz} J. Sollfrank and U. Heinz, in {\it Quark--Gluon
Plasma 2}, ed. R. C. Hwa
(World Scientific, Singapore,
1995).

\bibitem{becattini} F. Becattini and U. Heinz,
{\it Z. Phys.} C76 (1997) 269.

\bibitem{cleymans} J. Cleymans and K. Redlich,
{\it Phys. Rev. Lett.} 81 (1998) 5284.

\bibitem{bhs} P. Braun-Munzinger, I. Heppe and J. Stachel,
{\it Phys. Lett.} B465 (1999) 15.

\bibitem{zbsg} E. E. Zabrodin, L. V. Bravina,
H. St\"ocker and W. Greiner,
hep-ph/9901356 (1999).

\bibitem{fermi} E. Fermi,
{\it Prog. Theor. Phys.} 5 (1950) 570; {\it Phys. Rev.} 81 (1951)
683.

\bibitem{landau} L. D. Landau, {\it Izv. Akad. Nauk
Ser. Fiz.} 17 (1953) 51; S. Z. Belenkij and L. D. Landau, {\it
Nuovo Cim. Suppl.} 3 (1956) 15; I. J. Pomeranchuk, {\it Dokl. Akad. Nauk
Ser. Fiz.} 78 (1951) 889.

\bibitem{hagedorn} R. Hagedorn,
{\it Nuovo Cim. Suppl.} 3 (1965) 147.

\bibitem{blz} T. S. Bir\'o, P. L\'evai and J. Zim\'anyi,
{\it Phys. Rev.} C59 (1999) 1574; J. Zim\'anyi, T. S. Bir\'o, T.
Cs\"org\"o and P. L\'evai, hep-ph/9904501 (1999).

\bibitem{satzqm99} H. Satz, in
{\it
Proceedings of Quark Matter 99}
(Torino, Italy, May 10th-15th 1999), to appear in {\it Nucl. Phys.}
A (hep-ph/9908339, 1999).

\bibitem{isi} M. B. Isichenko, {\it Rev. Mod. Phys.} 64 (1992) 961;
D. Stauffer and A. Aharony, {\it Introduction to Percolation Theory}
(Taylor \& Francis, London, 1994).

\bibitem{dias} J. Dias de Deus, R. Ugoccioni and A. Rodrigues,
preprint FISIST-10-99-CENTRA
(hep-ph/9907352, 1999). 

\bibitem{baym} G. Baym, {\it Physica} 96A (1979) 131; T. Celik, F.
Karsch and H. Satz, {\it Phys. Lett.} B97 (1980) 128.

\bibitem{gaisser} T. K. Gaisser, {\it Cosmic Rays and Particle
Physics} (Cambridge University Press, Cambridge, 1991).

\bibitem{cosmic2} C. Pajares, D. Sousa and R. A. V\'azquez,
hep-ph/9805475 (1998), to appear in {\it Astropart. Phys.}

\bibitem{cosmic3} G. Battistoni, C. Forti, J.
Ranft and S. Roesler, {\it Astropart. Phys.} 7 (1997) 49; G.
Battistoni, M. Carboni, C. Forti and J. Ranft,
preprint INFN-AE-99-07 (1999).

\bibitem{cosmic4} D. Heck, G. Schatz, T. Thouw,
J. Knapp and J. N. Capdevielle, preprint
FZKA-6019 (1998); D. Heck and J. Knapp,
preprint FZKA-6097 (1998).

\bibitem{eskolacern} K. J. Eskola, in {\it Proceedings of
the International Europhysics Conference on High Energy Physics}
(Tampere, Finland, July 15th-21st 1999), to be published by IOP
Publishing (hep-ph/9911350, 1999).

\end{thebibliography}
\end{document}